\documentclass[a4paper,11pt]{article}
\usepackage{jcappub} % for details on the use of the package, please see the JINST-author-manual
\usepackage{lineno}
\usepackage{subfigure}
\usepackage{xcolor}
\usepackage{soul}
\usepackage{physics}
\usepackage{comment}
\usepackage{amsmath}
\usepackage{booktabs}

\arxivnumber{} % if you have one

\arxivnumber{2601.14413}

\title{\boldmath Amplifying the Cosmological Collider with Ghost Spectators}

\author[a]{Matheus~C.~Ferreira,}
\author[a,b]{Felipe~T.~Falciano,}
\author[c]{and Guilherme L. Pimentel}

\affiliation[a]{Brazilian Center for Research in Physics (CBPF), \\ Rua Dr. Xavier Sigaud 150, 22290-180, Rio de Janeiro, RJ, Brazil}
\affiliation[b]{PPGCosmo, CCE, Federal University of Espírito Santo (UFES), \\
Av. Fernando Ferrari 540, 29075-910, Vitória, ES, Brazil}
\affiliation[c]{Scuola Normale Superiore and INFN,\\ Piazza dei Cavalieri 7, 56126, Pisa, Italy}

\emailAdd{matheuscurado@cbpf.br}
\emailAdd{ftovar@cbpf.br}
\emailAdd{guilherme.leitepimentel@sns.it}

\abstract{Ghost inflation is a well-known framework in which cosmological fluctuations can generate enhanced primordial non-Gaussianity, typically of the equilateral type. In its original form, however, it is in tension with current observational constraints. Here we instead consider a setup in which a standard inflaton drives the background evolution, while excitations of a ghost condensate act as spectator fields that interact with the inflaton. This proposal fits naturally within the cosmological collider program: the exchanged particle has a modified dispersion relation, $\omega
\propto k^2$. We show that this ghost-inspired dynamics weakens the usual Boltzmann suppression, similarly to models with a very small effective sound speed, yielding
an enhanced bispectrum signal relative to standard cosmological collider scenarios. At the same time, the horizon-crossing scale remains a free parameter of the theory. As a result, the model shares features of both the de Sitter bootstrap and boostless
frameworks. Finally, we derive the differential equations governing cosmological correlators in the ghost-collider setup. Their structure reflects the quadratic momentum dependence of the dispersion relation and distinguishes this scenario from
conventional relativistic cases.}

\begin{document}
\maketitle
\flushbottom

%%%%%%%%%%%%%%%%%%%%%%%%%%%%%%%%%%
%%%%%%%%%%%%%%%%%%%%%%%%%%%%%%%%%%
%%%%%%%%%%%%%%%%%%%%%%%%%%%%%%%%%%
%%%%%%%%%%%%%%%%%%%%%%%%%%%%%%%%%%
\section{Introduction}
\label{sec:intro}

Primordial fluctuations encode, in their non-Gaussianities, the signatures of the underlying physics of inflation. In the past decade, this has become clearly articulated by the concept of ``Cosmological Collider physics"~\cite{
Arkani-Hamed:2015bza,
Arkani-Hamed:2018kmz,
Pinol_2023,
Qin_2023,
Qin:2025xct,
%Yin:2023jlv,
Aoki:2023wdc,
Aoki:2024uyi,
Jazayeri:2023xcj,
Jazayeri:2023kji,
Jazayeri:2022kjy,
DuasoPueyo:2023kyh,
Baumann_2012,
Baumann_2018,
Lee_2016,
Wu_2019,
Saito_2018,
Lu_2020,
Lu_2021,
Sou_2021,
Hook_2020,
Hook_20201,
Liu_2020,
Wang_2020,
Fan_2021,
Kumar_2020,
Hubisz:2024xnj,
Cui_2022,
Chua:2018dqh,
Chen_2018,
Chen_2017,
Chen_2012,
Chen:2009zp,
Kehagias_2017,
Gong_2013,
Noumi_2013}.
During inflation 
\cite{
Guimaraes:2020vfd,
Iacconi:2024hmg,
Guth:1980zm,
Mukhanov:1990me,
Linde:1981mu,
Maldacena:2002vr,Martin2014,Planck:2018jri}, the rapid expansion of spacetime stretches quantum fluctuations of massive fields, allowing them to imprint distinctive non-analytic signatures in primordial correlation functions, before they redshift away. These features, which are sharpest in the squeezed limit of the bispectrum and the collapsed limit of the trispectrum, carry spectroscopic information on the mass and spin of the exchanged particle.\footnote{The mass appears as a feature as the ratio of long and short momenta is dialed, while the spin is a feature in the angular dependence between long and short modes.} This makes inflation a unique high-energy laboratory, with the potential to uncover degrees of freedom at or above the Hubble scale that would otherwise remain inaccessible.\footnote{The Hubble scale during inflation is still unknown; detecting gravitational waves from inflation would allow us to determine this scale in Planck units. Typical models of inflation predict that it is very large; phenomenologically, the scale can be very low, as long as the universe is reheated before Big Bang nucleosynthesis.}

In slow-roll inflation and quasi-single-field inflation, the primordial non-Gaussianities tend to be very small. Moreover, for a heavy particle\footnote{Strictly speaking, a particle in the principal series of $SO(1,4)$ irreducible representations.} of mass $M$ on a de Sitter background, the curvature-induced production rate is suppressed by the standard Boltzmann factor $e^{-\pi \mu}$, with $\mu \sim m/H$.\footnote{The bispectrum of the curvature perturbation $\zeta$ due to the exchange of a massive particle takes the characteristic form in the squeezed limit $k_3\ll k_1\sim k_2$, with $\cos\theta\equiv\mathbf{k}_1.\mathbf{k}_3/(k_1 k_3)$,
\begin{equation}
\langle \zeta_{\mathbf{k}_1}\,\zeta_{\mathbf{k}_2}\,\zeta_{\mathbf{k}_3} \rangle'
\;\propto\;
\left[ e^{-\pi \mu }\left(\frac{k_3}{k_1}\right)^{\Delta_-} + (\Delta_-\leftrightarrow \Delta_+)\right] P_s(\cos \theta), 
\qquad
\Delta_{\pm} = -\frac{3}{2} \pm i\mu,
\qquad
\mu = \sqrt{\frac{m^2}{H^2} - \frac{9}{4}}. \nonumber
\end{equation}
When \( m > 3H/2 \), the parameter \( \mu \) becomes real, and the bispectrum develops an oscillatory dependence on the logarithm of the momentum ratio, 
\(\ln (k_3/k_1)\).
This logarithmic oscillation encodes the mass of the exchanged particle, producing an interference-like pattern in momentum space.} There are proposals to mitigate this suppression using non–Bunch–Davies vacua~\cite{Yin:2023jlv}, couplings where the inflaton speed acts as a chemical potential \cite{Bodas_2021,Bodas:2024hih}, or including additional interactions engineered to enhance the non-Gaussian signal \cite{Pimentel:2022fsc,Qin:2023ejc,Wang:2025qww}.

In this paper, we consider a model inspired by the ghost condensate, where Lorentz invariance is spontaneously broken~\cite{Hamed_2004, Nima_Arkani_Hamed_2004}. The corresponding scalar fluctuation exhibits a modified dispersion relation $\omega^2 \propto k^4$, characteristic of higher-derivative operators, which significantly changes particle production in the inflationary bulk. In particular, the late-time mode functions involve Hankel functions with argument $(k\eta)^2$ and index $\mu/2$, effectively softening the exponential suppression and allowing heavy ghost fields with $m \gg H$ to contribute to primordial correlators.
\begin{table}[ht]
\centering
\begin{tabular}{lccc}
\toprule
 & \textbf{dS} & \textbf{Boostless} & \textbf{Ghost} \\
\midrule
Horizon crossing 
& $k/a  \sim H$ 
& $ k/a \sim H/c_s $ 
& $ k/a \sim H/\sqrt{\gamma}$ \\
\\
Suppression in $f_{NL}$
& $e^{-\pi \mu}$ 
& $e^{-\pi \mu / 2} \lesssim f_n(c_s) \lesssim e^{-\pi \mu}$ 
& $e^{-\pi \mu / 2}$ \\
\\
Oscillatory phase 
& $\delta(\mu)$ 
& $\delta(\mu, c_s)$ 
& $\delta(\mu, \gamma)$ \\
\bottomrule
\end{tabular}
\caption{Comparison between de Sitter, boostless, and ghost condensate scenarios. While we have the sound horizon for the boostless case, we have a ``ghost" horizon for the ghost condensate.~The location of the ghost horizon is adjustable without affecting the amplitude of $f_{\rm NL}$, unlike the boostless case.}
\label{tab:ds_boostless_ghost}
\end{table}\\
In fact, massive ghost particles do not experience the same Boltzmann suppression as their de Sitter counterparts, resulting in less suppression in both the bispectrum and trispectrum.\footnote{Despite the name ``ghost condensate," the theory is perfectly unitary. For a recent paper considering the effects of ``standard" ghosts in the cosmological collider, see \cite{Aoki:2025uff}.} Our analysis shows that, for sufficiently large masses, the enhancement can reach several orders of magnitude. This renders the ghost-mediated signal observationally relevant even in mass ranges that would otherwise be entirely inaccessible. At the same time, the boundary behavior of the correlators remains controlled by scale (but not conformal) symmetry.

To make these statements precise, we construct the relevant bulk-to-bulk and bulk-to-boundary propagators within the Schwinger–Keldysh formalism \cite{Chen:2017ryl, Weinberg_2005, calzeta}. We derive the mixed case where inflaton modes interact through the exchange of a massive ghost. The higher-derivative dispersion relation is such that the seed functions have different conformal weight---instead of $\Delta = 2, 1$  in the de Sitter invariant case, they have $\Delta = \tfrac52, \tfrac12$ for the ghost. As a consequence of exchanging this ``ghost," the leading non-analytic contributions persist, albeit with modified scaling behavior and substantially reduced exponential suppression. In the collapsed limit, we demonstrate explicitly that the characteristic oscillatory features remain intact with smaller Boltzmann suppression.

After discussing the cosmological collider phenomenology, we derive the bootstrap equations corresponding to the ``ghost collider" scenario, showing that they contain explicit higher-derivative contributions that are absent in the standard de Sitter case. These equations define a generalized hypergeometric system with a modified singularity structure, while still obeying the boundary constraint imposed by the residual conformal symmetry. The emergence of higher-derivative terms in the bootstrap operator suggests a broader program of exploring bootstrap equations for general dispersion relations of exchanged particles. 
\vskip 4 pt
\noindent {\bf Outline}: This paper is organized as follows. In Section 2, we review the basic ingredients of field theory in de Sitter space. Section 3 introduces the ghost spectator particle and discusses how its dynamics differ from those of standard de Sitter fields. In Section 4, we develop the cosmological collider framework for this setup and present the computation of the trispectrum using the Schwinger–Keldysh formalism. Section 5 is devoted to the analysis of the ghost-mediated bispectrum and its phenomenological implications. In Section 6, we derive the modified bootstrap equations associated with the ghost collider scenario and study their singularity structure. Finally, Section 7 summarizes our main results and discusses possible directions for future theoretical and observational studies.

\noindent {\bf Notation and Conventions}: We work with the metric signature \( (-,+,+,+) \). Throughout the paper, we use natural units,
\( \hbar = c = 1 \). Scalar fields associated with the inflaton sector are denoted by
\( \varphi \) and \( \phi \), while ghost fields are denoted by \( \sigma \).
Spacetime indices are labeled by \( \mu = (\eta, 1, 2, 3) \), where \( \eta \) denotes conformal
time, which is used throughout this work. Latin indices \( i = 1,2,3 \) refer to spatial
components. Three dimensional vectors are written in boldface, \( \mathbf{k} \). The magnitude of a vector
is defined as \( k = |\mathbf{k}| \), and the corresponding unit vector is
\( \hat{\mathbf{k}} = \mathbf{k}/k \). The momentum of the \( n \)-th external leg of a correlation
function is denoted by \( \mathbf{k}_{n} \), with magnitude \( k_{n} \equiv |\mathbf{k}_{n}| \).

\clearpage

%%%%%%%%%%%%%%%%%%%%%%%%%%%%%%%%%%
%%%%%%%%%%%%%%%%%%%%%%%%%%%%%%%%%%
%%%%%%%%%%%%%%%%%%%%%%%%%%%%%%%%%%
%%%%%%%%%%%%%%%%%%%%%%%%%%%%%%%%%%
\pagebreak
\section{Field Theory in dS} \label{sec:model}

In this section we establish the underlying cosmological scenario, contrasting with quantum field theory in de Sitter (dS) spacetime; the case where all isometries are preserved. 
Consider a scalar field $\phi$ with mass $m$ on a dS background. We use flat slicing with Euclidean coordinates $\vec{x}$ and conformal time $\eta$, where the line element reads
\begin{equation}
    \dd s^2=a(\eta)^2\Big(-\dd \eta^2+\dd \vec{x}\cdot\dd \vec{x}\Big)\ , \quad a(\eta)=-\frac{1}{H\eta} \ ,
\end{equation}
and the Hubble factor $H=\partial_{\eta}a/a^2 $ is a constant. The equation of motion of a Fourier mode $\phi(k,\eta)$ is
\begin{equation}\label{motionbackground}
    \partial_{\eta}^2\phi(k,\eta)-\frac{2}{\eta}\partial_{\eta}\phi(k,\eta)+\bigg( k^2 + \frac{m^2}{\eta^2 H^2}\bigg) \phi(k,\eta) = 0 \ .
\end{equation}
Using standard canonical quantization, the mode function is 
\begin{align}
   u_k(\eta) &=\frac{H\sqrt{\pi}}{2}e^{i \pi / 4}e^{- \pi \mu / 2} (-\eta)^{3/2} \mathcal{H}_{i\mu}^{(1)}(-k\eta)\ , \label{dsmodeHankel}
\end{align}
where $\mathcal{H}^{(1)}_{i\mu}$ is the Hankel function of the first kind, and the coefficient is given by
\begin{equation}\label{hankelcoeficient}
\mu = \sqrt{m^2/H^2-9/4}\ .
\end{equation}
This solution picks the Bunch-Davies vacuum $b_{k}|0\rangle = 0$, that is, in the far past the mode function reproduces the standard flat spacetime solution
\begin{equation}
    \lim_{\eta \to -\infty} u_{k}(\eta) = i H \eta\frac{e^{-ik\eta}}{\sqrt{2k}} \ .
\end{equation}
In the asymptotic future, the massive scalar field behaves as
\begin{equation}\label{anomalousdimen}
    \lim_{\eta \to 0} \phi(k,\eta) \propto O^{+}(k)\eta^{\Delta_{+}}+O^{-}(k)\eta^{\Delta_-} \ ,
\end{equation}
with the conformal dimensions given by $\Delta_{\pm} = \frac{3}{2} \pm i\mu$. In the small-argument expansion of the two-point correlation function
\begin{align}\label{desitterszero}
    & \lim_{k \to 0}\langle  \phi_{k}(\eta_1) \phi_{-k}(\eta_2) \rangle  \propto  \frac{2}{\pi k^3}H^2 \bigg[ \Gamma(-i\mu)^2 \bigg( \frac{k^2 \eta_1 \eta_2}{4} \bigg)^{\Delta_+}+ \Gamma(i \mu)\bigg( \frac{k^2 \eta_1 \eta_2}{4} \bigg)^{\Delta_{-}}  \bigg] \ .
\end{align}

In Minkowski spacetime, the expectation value decays exponentially. The oscillatory behavior in Eq.~\eqref{desitterszero}, arising from the imaginary anomalous dimension, signals the creation of particle-antiparticle pairs driven by the expansion of the universe~\cite{Arkani-Hamed:2015bza}. This happens only for heavy particles, since for light or massless particles, $\mu$ is imaginary and $\Delta_\pm$ are real. Although the two-point function is non-zero for any mass, heavy particles have small correlators, being exponentially suppressed by a Boltzmann factor $e^{-\pi \mu}$. Notwithstanding, heavy scalar fields can still be detected indirectly through interference effects, analogous to those observed in the double-slit experiment. When a massive particle couples to a massless scalar—such as the inflaton—it imprints a distinctive signal on higher-order correlation functions. By analyzing the bispectrum and trispectrum of the massless field, one can extract information about the properties of the heavy particle.

%%%%%%%%%%%%%%%%%%%%%%%%%%%%%%%%%%
%%%%%%%%%%%%%%%%%%%%%%%%%%%%%%%%%%
%%%%%%%%%%%%%%%%%%%%%%%%%%%%%%%%%%
%%%%%%%%%%%%%%%%%%%%%%%%%%%%%%%%%%

\section{The Cosmological Model}\label{section:ghostspectator}

In our model, gravitational dynamics is governed by the Einstein--Hilbert action. The inflaton drives the cosmological evolution of the Universe, whereas the ghost field is introduced as a spectator scalar that interacts with the inflaton sector. The dynamics of the system are described by the action
\begin{equation}
S = S_{\text{gravity}} - \int d^{4}x \,
\big( \mathcal{L}_{\phi} + \mathcal{L}_{\sigma} + \mathcal{L}_{\text{int}} \big),
\label{eq:total_action}
\end{equation}
where \( \mathcal{L}_{\phi} \), \( \mathcal{L}_{\sigma} \), and \( \mathcal{L}_{\text{int}} \) denote the Lagrangian density of the inflaton, the ghost spectator field, and their interaction, respectively. The inflaton sector is assumed to take the standard single-field inflationary form,
\begin{equation}
\mathcal{L}_{\phi}
= %\left[
-\frac{1}{2}(a \partial_{\eta} \phi )^2 +\frac{1}{2}(a \nabla \phi )^2 
+ a^4V(\phi)
%\right]
\quad ,
\label{eq:inflaton_lagrangian}
\end{equation}
where \( V(\phi) \) is the inflaton potential responsible for driving inflation. The spectator Lagrangian for the massive ghost is the following \cite{Hamed_2004,Nima_Arkani_Hamed_2004},
\begin{equation}\label{L2sigmaaction}
\mathcal{L}_{\sigma} = \frac{M^4}{2}\Big\{-(a \, \partial_\eta\sigma)^2 + \frac{1}{M^2}(\nabla^2 \sigma)^2  + a^4 m^2_{ghost}\,\sigma^2 \Big\}\ .
\end{equation}

We assume that the ghost is much heavier than the inflaton, but its mass must be smaller than the cutoff scale $(m_{ghost}\ll M)$. We can give a rough estimate of the mass of the ghost particle by comparing it with the inflaton mass. The typical inflation energy scale gives a Hubble factor of the order of $H \sim 10^{13} \ \text{GeV} \sim 10^{-5}\, m_{pl} $, which gives a lower bound for the cutoff mass since $H < M $. During a single-field slow roll, the inflaton mass $m^2_{inflaton} \sim \eta_{V} H^2$, where $\eta_{V}\sim 10^{-2}$ is the dimensionless slow roll parameter $ \eta_V \equiv \partial_{\phi \phi }V /V$~\cite{Baumann:2022mni,Planck:2018jri}. A conservative estimate is to consider the cutoff mass below the Planck scale, say $M\sim10^{-2}\, m_{pl} $; therefore, the range for the mass of the ghost particle is
\begin{equation}
    10^{-1} < \frac{m_{gh}}{H} < 10^3 \ .
\end{equation}

%%%%%%%%%%%%%%%%%%%%%%%%%%%%%%%%%%
%%%%%%%%%%%%%%%%%%%%%%%%%%%%%%%%%%
%%%%%%%%%%%%%%%%%%%%%%%%%%%%%%%%%%
%%%%%%%%%%%%%%%%%%%%%%%%%%%%%%%%%%
\subsection{Mode functions for Ghosts and Inflatons} \label{sec:modes-and-powerspec}
In the next section, we will determine the four-point correlation functions for different scenarios. The basic input of this construction is the mode function that encodes the evolution of each particle. There are three main scenarios of interest: a massless inflaton particle, a massless ghost particle, and a massive ghost particle.\footnote{Let us stress again that ``ghost" here means an excitation of the ghost condensate; nonetheless they have positive kinetic terms and no issues with negative probabilities.}~First, consider the excitation of an inflaton particle described by the scalar quantum field $\varphi(\vec{x},\eta)$,. Its Fourier modes can be written in terms of creation and annihilation operators as
\begin{equation}
\varphi_k(\eta)=u(k,\eta)b_{\bf k}+u^\ast (k,\eta)b^\dagger_{-\bf k} \quad ,
\end{equation}
where $b_{\bf k}\ket{0}=0$ and $u(k,\eta)$ is chosen {\it a la} Bunch-Davies.~At zeroth order in slow-roll, inflation reduces to de Sitter space \cite{Baumann:2022mni, Creminelli:2012qr,Mukhanov:1990me, Linde:1981mu, Maldacena:2002vr}, and the equations of motion read
\begin{equation}\label{infmodedyn}
\left[\partial_{\eta}^2 -\frac{2}{\eta} \partial_{\eta}+ k^2+\frac{m^2}{(H \eta)^2}
\right]u(k,\eta) = 0 \ .
\end{equation}
The solution is a linear combination of two Hankel functions. In the far past, $\eta \to -\infty$, we expect to have a flat spacetime behavior $ u(k,\eta) \propto \ e^{-ik\eta} $, which select only the Hankel function of the first kind $\mathcal{H}^{(1)}_{\nu}$. Therefore, the solution of Eq.~\eqref{infmodedyn} is
\begin{align}
u(k,\eta) &=\frac{H\sqrt{\pi}}{2\, k^{3/2}}e^{i \pi (1+2i\mu)/ 4} (-k\eta)^{3/2} \mathcal{H}_{i\mu}^{(1)}(-k\eta)
\hspace{0.5cm}\text{(generic inflaton field)}\ ,
\label{modeinfm}
\end{align}
where the parameter $\mu=\sqrt{m^2/H^2-9/4}$. There are two important cases for our study: conformally coupled  particles with $m^2=2H^2$, and massless inflaton. The mode functions are, respectively, given by
\begin{equation}
u(k,\eta) = \begin{cases}
-\frac{H\eta}{\sqrt{2k}}e^{-ik\eta}, & \text{(cc scalar; $m^2=2H^2$)} \\
\frac{-H\eta}{\sqrt{2 k}}\bigg(1-\frac{i}{k \eta}\bigg)e^{-ik\eta}, &\text{(inflaton; $m=0$)} \label{modeinfm0}\end{cases}
\end{equation}

The ghost field follows a similar description, being a scalar quantum field $\sigma(\vec{x},\eta)$ whose Fourier mode can be decomposed as
\begin{equation}
\sigma_k(\eta)=v(k,\eta)B_{\bf k}+v^\ast (k,\eta)B^\dagger_{-\bf k} \quad ,
\end{equation}
where the annihilation operator $B_{\bf k}$ now annihilates a different vacuum $B_{\bf k}\ket{0_{ghost}}=0$. The dynamics of the ghost spectator follows from Eq.~\eqref{L2sigmaaction}, so the equation of motion is
\begin{equation}\label{ghostdynamicsbulk}
    \left[\partial_{\eta}^2  -\frac{2}{\eta}\partial_{\eta} + \bigg(\gamma^2 k^4 \eta^2 + \frac{m^2_{\text{ghost}}H^{-2}}{\eta^2 }\bigg)\right]v(k,\eta) = 0 \ ,
\end{equation}
where we defined the parameter $\gamma \equiv H/M$. The solution gives the mode function of the ghost particle. The general solution of \eqref{ghostdynamicsbulk} is a combination of Hankel functions but with the argument given by $k^2 \eta^2$. In the asymptotic limit $\eta \to -\infty$, the field oscillates with $e^{\pm i k^2 \eta^2}$. The initial vacuum selects only one mode. Therefore, we obtain
\begin{align} 
\label{sigmamotion}
v(k,\eta) &=
H\sqrt{\frac{\pi}{8}}e^{-\pi (i+\mu)/4}(-\eta)^{3/2}\mathcal{H}^{(1)}_{ i\mu/2}\left( \frac{\gamma k^2 \eta^2}{2} \right)
&&
\text{(generic ghost field)} \\
&=
-\frac{H}{k}\sqrt{\frac{\eta}{2\gamma}}e^{i\gamma k^2 \eta^2/2}
&&
\text{(ghost; $m^2=5H^2/4$)} \label{ghostm5H4}\\
&=
H\sqrt{\frac{\pi}{8}}e^{i \pi/ 8}(-\eta)^{3/2}\mathcal{H}^{(1)}_{ 3/4}\left( \frac{\gamma k^2 \eta^2}{2} \right)
&&
\text{(ghost; $m=0$)} \label{ghostm0}
\end{align}
where again $\mu = \sqrt{m^2_{\text{ghost}}/H^2-9/4}$ but with the mass of the ghost particle. Although the mode function closely resembles the usual inflationary scenario, the modified dispersion relation introduces significant differences. Specifically, the argument for the Bessel function becomes quadratic in $k\eta$, and the index is half compared to the conventional case. In the large-scale limit, the two-point function behaves as follows:
\begin{align}\label{ghostszero}
    & \lim_{k\to 0} \langle  \sigma_{k}(\eta_1) \sigma_{-k}(\eta_2) \rangle  \propto  \frac{H^2}{\pi k^3}\bigg[ \Gamma(-i\mu/2)^2 \gamma^{i \mu} \bigg( \frac{ k^2 \eta_1 \eta_2}{4} \bigg)^{\Delta_+}+ \text{c.c} \bigg],
\end{align}

As noted previously, we have an oscillatory behavior, but this is less strongly suppressed due to the additional term $\gamma$ and the change in the argument of the Gamma functions, which now is $\pm i\mu/2$.~These preliminary analyses provide a basis for distinguishing the non-Gaussian signal from the standard de Sitter case.

%%%%%%%%%%%%%%%%%%%%%%%%%%%%%%%%%%
%%%%%%%%%%%%%%%%%%%%%%%%%%%%%%%%%%
%%%%%%%%%%%%%%%%%%%%%%%%%%%%%%%%%%
%%%%%%%%%%%%%%%%%%%%%%%%%%%%%%%%%%

\section{Cosmological Collider} \label{sec:collider}

A fundamental idea behind the cosmological collider framework is that interactions between particles in the bulk leave imprints on the cosmological correlators at the boundary. Moreover, the correlators can be reduced to a unique building block, the four-point function of conformally coupled scalars, arising from the exchange of a massive scalar. The seed function typically serves as the starting point for constructing the full set of cosmological correlators.

We consider the cubic interaction $\mathcal{L}_{3,\text{int}} \supset \varphi^2 \sigma $, where it is assumed that $\varphi$ is a conformally coupled field, that is, it has a scaling dimension ${\Delta=2}$. In order to obtain the correlators of massless external fields, for which the scaling dimensions ${\Delta=3}$, one can apply the weight-shifting operator. Another interesting interaction is $\mathcal{L}_{3,\text{int}} \supset (\partial_{\eta}\varphi)^2 \sigma $ that emerges in quasi-single-field models~\cite{Chen:2009zp,Chen_2012,Noumi_2013,Chen:2017ryl,Baumann_2018}. It generates a four-point function with four massless external fields exchanging a massive field in the bulk, which is in the core of Cosmological Collider Phenomenology
\cite{Arkani-Hamed:2015bza,
Arkani-Hamed:2018kmz,
Pinol_2023,
Qin:2025xct,
%Yin:2023jlv,
Aoki:2024uyi,
Jazayeri:2023xcj,
Jazayeri:2023kji,
Jazayeri:2022kjy,
DuasoPueyo:2023kyh,
Baumann_2012,
Baumann_2018,
Lee_2016,
Wu_2019,
Lu_2020,
Lu_2021,
Hook_2020,
Hook_20201,
Liu_2020,
Wang_2020,
Kumar_2020,
Hubisz:2024xnj,
Cui_2022,
Chua:2018dqh,
Chen_2018,
Chen_2017}.

%%%%%%%%%%%%%%%%%%%%%%%%%%%%%%%%%%
%%%%%%%%%%%%%%%%%%%%%%%%%%%%%%%%%%
%%%%%%%%%%%%%%%%%%%%%%%%%%%%%%%%%%
\subsection{Basic formalism}

The standard framework for computing cosmological correlators in the presence of external perturbations is the \emph{in--in} formalism or the Keldysh--Schwinger formalism~\cite{Weinberg_2005,Chen:2017ryl,calzeta,Chen_2010}, and reviewed in the Appendix~\ref{appendixkeldysh}. Here, we summarize only the essential results to fix the notation. Consider a scalar field decomposed into a homogeneous background and a perturbation,
\[
\phi(\eta, \vec{x}) = \phi_0(\eta) + \varphi(\eta, \vec{x}) \, .
\]
This separation allows us to treat the background field as a fixed, time-dependent quantity, while the perturbations are promoted to quantum fields. Accordingly, the Lagrangian density is expressed as a functional of the fluctuations.

As in standard quantum field theory, the $n$-point correlation function is obtained by differentiating the partition function with respect to the external source. In our case, we are interested in the interaction between two inflaton fields mediated by a massive particle. The ghost field propagates in the bulk, whereas only the inflaton perturbations survive to the boundary. It is therefore convenient to introduce two types of propagators: one connecting points within the bulk and another linking a bulk point to the boundary. The bulk-to-bulk propagator encodes the dynamics of the massive particle.
\begin{align}
&G_{++}(k,\eta_1,\eta_2) = \  \sigma_{k} (\eta_1) \sigma_{k}^* (\eta_2) \Theta (\eta_1 - \eta_2)+ \sigma_{k}^*(\eta_1)\sigma_{k}(\eta_2) \Theta (\eta_2 - \eta_1)\ ,\nonumber\\
&G_{--}(k,\eta_1,\eta_2) = \ \sigma_{k}(\eta_1)\sigma_{k}^*(\eta_2) \Theta (\eta_2 - \eta_1)+\sigma_{k}^*(\eta_1)\sigma_{k}(\eta_2) \Theta (\eta_1 - \eta_2)\ , 
\label{diafg}\\
&G_{+-}(k,\eta_1,\eta_2) = \sigma_{k}^*(\eta_1)\sigma_{k}(\eta_2)\ , \nonumber\\
&G_{-+}(k,\eta_1,\eta_2) = \sigma_{k}(\eta_1)\sigma_{k}^*(\eta_2)\ .\label{nondifr}
\end{align}

The function $\Theta(\eta)$ denotes the Heaviside step function. In the \emph{in--in} formalism, the plus sign corresponds to the time-ordered contribution, whereas the minus sign represents the anti-time-ordered piece along the integration contour. The bulk-to-bulk propagators satisfy the complex-conjugate relations
\[
G_{++}(k,\eta_1,\eta_2) = G^{\ast}_{--}(k,\eta_1,\eta_2)
\quad \text{and} \quad
G_{+-}(k,\eta_1,\eta_2) = G^{\ast}_{-+}(k,\eta_1,\eta_2) \, .
\]
Massless particles propagate from the bulk to the boundary, defined by the limit $\eta_0 \rightarrow 0$. Accordingly, we define the bulk-to-boundary propagators as
\begin{align}\label{btob}
K_{+}(k,\eta) &= \varphi_{k}(\eta_0)\,\varphi_{k}^{\ast}(\eta) \ , 
&\qquad 
K_{-}(k,\eta) &= \varphi_{k}^{\ast}(\eta_0)\,\varphi_{k}(\eta) \ ,
\end{align}
where the subscripts have the same meaning as before, and $K_{+}(k,\eta) = K_{-}^{\ast}(k,\eta)$. In general, Eqs.~\eqref{diafg}--\eqref{nondifr} define the four distinct types of bulk-to-bulk propagators, while Eq.~\eqref{btob} defines the two types of bulk-to-boundary propagators. Rather than computing all possible diagrams, we can focus on a single diagram that captures the phenomenologically relevant non-Gaussian signal: the four-point interaction diagram with a scalar exchange in the bulk. This approach substantially simplifies the analysis, as it captures the dominant non-Gaussian features that appear in the power spectrum and higher-order correlation functions. To construct such diagrams, we follow a systematic set of rules governing the arrangement of propagators and interaction vertices within the \emph{in--in} contour \cite{Chen:2017ryl}.

%%%%%%%%%%%%%%%%%%%%%%%%%%%%%%%%%%
%%%%%%%%%%%%%%%%%%%%%%%%%%%%%%%%%%
%%%%%%%%%%%%%%%%%%%%%%%%%%%%%%%%%%
\subsection{Trispectrum and the Seed Function}
It is straightforward to express the four-point correlation function associated with the interaction term $\varphi^2\sigma$. Using the Schwinger-Keldysh formalism, the four-point function is given by the integral
\begin{align}
\left<\varphi_{\bf k_1}\varphi_{\bf k_2}\varphi_{\bf k_3} \varphi_{\bf k_4}
\right>^{\prime}_\sigma
&=4 (-ig)^2\sum_{a,b = \pm}ab \int \frac{d\eta_1}{(-H\eta_1)^4} \frac{d\eta_2}{(-H\eta_2)^4}  \ K_{a}(k_1,\eta_1) K_{a}(k_2,\eta_1) \nonumber \\
    & \times  G_{ab}(s,\eta_1,\eta_2) K_b(k_3,\eta_2) K_b(k_4,\eta_2) +  \text{permutations} \ ,\label{fourpointconformalds}
\end{align}
where $\textbf{s} \equiv \textbf{k}_1 +\textbf{k}_2 = - \textbf{k}_3-\textbf{k}_4$. The permutation is taken over all external momenta, and the indices $a,b$ label the signs $\pm$. The quantities $k_1,k_2,k_3,k_4$ denote the magnitudes of the corresponding momenta, e.g. $k_1 \equiv | \textbf{k}_1|$. The subscript $\sigma$ specifies the nature of the exchanged particle in the bulk, whereas the prime on the correlator indicates that it is the stripped correlator, that is, the overall Dirac delta function that enforces momentum conservation has been removed.\footnote{Explicitly, it is defined as
\begin{equation}
\left<\varphi_{\bf k_1}\varphi_{\bf k_2}\varphi_{\bf k_3} \varphi_{\bf k_4}
\right>_\sigma
\equiv 
(2\pi)^3 \delta^{(3)}(\textbf{k}_1+\textbf{k}_2+\textbf{k}_3+\textbf{k}_4)
\left<\varphi_{\bf k_1}\varphi_{\bf k_2}\varphi_{\bf k_3} \varphi_{\bf k_4}
\right>^{\prime}_\sigma 
 \ .
\end{equation}}

We have one propagator $G_{ab}(s,\eta_1,\eta_2)$ representing the exchange of a massive particle in the bulk, and four bulk-to-boundary propagators $K_a(k,\eta)$, one for each inflaton mode that persists to the boundary. It is also straightforward to construct the four-point function corresponding to the interaction term $(\partial_\eta \varphi)^2\sigma$, which is characteristic of quasi-single-field inflationary models. The only modification with respect to the previous case is that the time derivative in the interaction acts on the propagator of the associated field, meaning that each bulk-to-boundary propagator acquires an additional time derivative. In this case, the four-point function is given by
\begin{align}
\left<\varphi_{\bf k_1}\varphi_{\bf k_2}\varphi_{\bf k_3} \varphi_{\bf k_4}
\right>^{\prime}_\sigma 
&=4 (-ig)^2\sum_{a,b = \pm}ab \int \frac{d\eta_1}{(-H\eta_1)^2} \frac{d\eta_2}{(-H\eta_2)^2} \ \partial_{\eta} K_{a}(k_1,\eta_1) \partial_{\eta}K_{a}(k_2,\eta_1)\nonumber \\
    &\quad \times  G_{ab}(s,\eta_1,\eta_2)\,  \partial_{\eta}K_b(k_3,\eta_2)\partial_{\eta}K_b(k_4,\eta_2) + \text{permutations},\label{fourpointinfla}
\end{align}

We now introduce the physical content by specifying the nature of the interacting particles. The dynamics of each field are encoded in their respective mode functions, which in turn determine the form of the propagators. The mode function of massless particles satisfies the same equation as a canonical scalar field in de~Sitter spacetime. We begin by constructing the four-point correlation function for a scenario involving only inflaton fields (massless particles). Sequently, we consider a mixed configuration in which the inflaton fields interact through the exchange of a massive ghost particle.

%%%%%%%%%%%%%%%%%%%%%%%%%%%%%%%%%%
%%%%%%%%%%%%%%%%%%%%%%%%%%%%%%%%%%
%%%%%%%%%%%%%%%%%%%%%%%%%%%%%%%%%%
%%%%%%%%%%%%%%%%%%%%%%%%%%%%%%%%%%
\subsection*{Pure de Sitter Seed Function}

The previously computed correlators share a common structure that admits a closed-form expression in the context of cosmic inflation: the so-called seed function~\cite{Qin:2023ejc,Baumann:2019oyu}. Since our goal is to evaluate the four-point function by deforming the de~Sitter theory, we require the bulk-to-boundary and bulk-to-bulk propagators, $K$ and $G$, corresponding to a field theory in de~Sitter spacetime. For the interaction term $\mathcal{L}_{3,\text{int}} \supset \varphi^2\sigma$, the fields $\varphi$ are conformally coupled and interact through the exchange of a heavy field $\sigma$. The conformally coupled fields are characterized by the scaling dimension $\Delta_{\varphi} = 2$, which corresponds to a mass $m^2 = 2H^2$, leading to the parameter $i\mu = 1/2$. The bulk-to-boundary propagators are constructed from the mode function given in Eq.~\eqref{modeinfm0}, and therefore we obtain
\begin{align}
K_\pm(k,\eta)=   \frac{H^2\eta\, \eta_0}{2k}e^{\pm ik(\eta-\eta_0)}
\ ,
\end{align}
then, the integral Eq.~(\ref{fourpointconformalds}), becomes
\begin{align}
\left<\varphi_{\bf k_1}\varphi_{\bf k_2}\varphi_{\bf k_3} \varphi_{\bf k_4}
\right>^{\prime}_\sigma=\frac{-g^2 \eta_0^4}{4 k_1k_2k_3k_4 }\sum_{a,b = \pm} ab
&\int_{-\infty}^{0}\frac{d \eta_1}{\eta_1^2}e^{aik_{12}\eta_1} \int_{-\infty}^{0}\frac{d\eta_2}{\eta_2^2}e^{bik_{34}\eta_2}
\notag\\
&\times G_{ab}(s,\eta_1,\eta_2)
    + \text{permutations} \ ,\label{correfourconf}
\end{align}
where $k_{12} \equiv k_1 + k_2$ and  $k_{34} \equiv k_3 + k_4$. The interaction term $\mathcal{L}_{3,\text{int}} \supset (\partial_\eta\varphi)^2 \sigma$ has a closely related structure. In this case, we consider massless inflaton fields, whose mode function is given by Eq.~\eqref{modeinfm0}. Consequently, the time derivative acts on the bulk-to-boundary propagator, which now takes the form
\begin{align}
\partial_\eta K_\pm(k,\eta)&=
\frac{H^2\eta}{2k}
\left(1\pm i k\eta_0 \right)
e^{\pm ik(\eta-\eta_0)}
\ ,
\end{align}
and then the integral Eq.~\eqref{fourpointinfla} can be recast as
\begin{align}
\left<\varphi_{\bf k_1}\varphi_{\bf k_2}\varphi_{\bf k_3} \varphi_{\bf k_4}
\right>^{\prime}_\sigma 
= \frac{-g^2}{4 k_1 k_2 k_3 k_4 }\sum_{a,b = \pm}ab & \int_{- \infty}^{0} d\eta_1 e^{aik_{12}\eta_1} \int_{- \infty}^{0}d\eta_2e^{bik_{34}\eta_2}\notag\\
&\times G_{ab}(s,\eta_1,\eta_2) 
+\text{permutations}\ .\label{correfourinfla}
\end{align}

The correlation functions in Eq.~\eqref{correfourconf} and Eq.~\eqref{correfourinfla} exhibit the same underlying mathematical structure, differing only in the power of the time-dependent factors within the integral. More generally, we define the seed functions that encode the tree level inflationary correlators and takes the general form
\begin{align}\label{Seedgeral}
    \hat{\mathcal{I}}^{\,l_1,l_2}_{ab} 
    \equiv 
    -ab\, s^{5+l_1+l_2}
    \int_{-\infty}^{0} d\eta_1\, d\eta_2\, 
    (-\eta_1)^{l_1}(-\eta_2)^{l_2}
    e^{ai k_{12}\eta_1 + bi k_{34} \eta_2}
    G_{ab}(s,\eta_1,\eta_2) \; .
\end{align}

By choosing the appropriate values of the parameters $l_1$ and $l_2$, one recovers the desired scenarios. For instance, for the conformally coupled case with $\Delta = 2$, we have $l_1 = l_2 = -2$, whereas for the massless case with $\Delta = 3$, we find $l_1 = l_2 = 0$. The hat in the above expression indicates normalization: for each choice of $l_i$, this quantity becomes independent of the variable $s$. Although one could have introduced the seed function directly, the construction presented above clarifies its physical meaning and motivates its use. The expression in Eq.~\eqref{Seedgeral} assumes that the propagators are given by Hankel functions; however, an analogous formulation can be obtained for mode functions expressed in terms of Whittaker functions~\cite{Qin:2023ejc, Qin_2023, Aoki:2023wdc}.

%%%%%%%%%%%%%%%%%%%%%%%%%%%%%%%%
%%%%%%%%%%%%%%%%%%%%%%%%%%%%%%%%
%%%%%%%%%%%%%%%%%%%%%%%%%%%%%%%%
\subsection*{Ghost Collider: Seed Function for the Mixed Scenario}

In the previous section, we considered the pure de~Sitter case, in which two inflaton fields exchange a conventional massive particle. We now extend the discussion to a mixed configuration involving both inflaton and ghost fields. From a phenomenological perspective, this case is particularly appealing, as it may lead to distinctive observational signatures; however, it is also technically more involved.

The ghost scenario explicitly breaks Lorentz invariance \cite{Hamed_2004, Nima_Arkani_Hamed_2004}, raising the question of to what extent it can preserve the de~Sitter symmetries at the boundary that are common to standard inflationary models. The exchange of a massive ghost particle in the bulk provides an interesting framework to explore the interplay between the symmetry structure of the bulk theory and the resulting features of correlation functions at the inflationary boundary. For the interaction term $\varphi^2\sigma$, where $\sigma$ denotes a ghost excitation, the corresponding seed function is given by the general integral Eq.~\eqref{Seedgeral}, with $G_{ab}(s,\eta_1,\eta_2)$ representing the ghost propagator. Taking the limit $s\to0$ of the bulk propagator (see Appendix \ref{appendix::besselfunctions})
\begin{align}
\lim_{s\to0}s^3 G_{ab}(s,\eta_1,\eta_2)
&=
\frac{H^2}{\pi}
\bigg\{
(\eta_1\eta_2)^{\Delta_-}
\bigg(\frac{s^{2}}{4}\bigg)^{\Delta_-}
\gamma^{-i\mu}\,
\Gamma\!\left(\frac{i\mu}{2}\right)^2
\nonumber\\
&\qquad\qquad
+
\frac{2\pi}{\mu}
\bigg(\frac{s}{2}\bigg)^{\Delta_+ + \Delta_-}
(-\eta_1)^{\Delta_+}(-\eta_2)^{\Delta_-}
e^{\pi\mu/2}\,
\mathrm{csch}\!\left(\frac{\pi\mu}{2}\right)
+\text{c.c.}
\bigg\} \, .
\label{propghost}
\end{align}
At small exchanged momentum, the propagator contains both the non-analytic
terms proportional to $s^{2\Delta_\pm}$ and an analytic cross term of order
$s^{\Delta_+ + \Delta_-}=s^3$. The latter gives rise to a local contribution
to the bispectrum once the three-point limit is taken, whereas the former
encode the oscillatory non-analytic cosmological-collider signal\footnote{Although the analytic term is local in the exchanged soft momentum, it
should be distinguished from the genuinely local EFT background, since it
may still generate oscillatory behavior, even though it does not carry the
$\gamma$-dependence emphasized in the first contribution.}. Since all
Schwinger--Keldysh propagators share the same small-$s$ structure at this
order, there is no need to specify the indices $a,b$ explicitly here. In the
following, when deriving the seed function relevant for the oscillatory signal,
we keep only the non-analytic part.~As we saw in the previous section, the Schwinger-Keldysh formalism computes the expectation values instead of in-out amplitudes.~Taking all the contributions, the late time expectation value reads
\begin{equation}
\mathcal{\hat{I}}^{l_1,l_2}\equiv \mathcal{\hat{I}}^{l_1,l_2}_{++} + \mathcal{\hat{I}}^{l_1,l_2}_{--} + \mathcal{\hat{I}}^{l_1,l_2}_{+-} + \mathcal{\hat{I}}^{l_1,l_2}_{-+}
\end{equation}
After expanding the modes for $s\to 0$, taking their leading contribution and replacing it in the propagators, the only remaining integral is of the form
\begin{align}
    J^{l,\Delta}_{\pm}(k)\equiv \pm i\int_{-\infty }^0  (-\eta)^l (-\eta)^{\Delta} e^{\pm i k \eta  } \, d\eta   = \pm i e^{ \mp i\frac{\pi}{2} (l+\Delta)}k^{-(1+l+\Delta )}\Gamma  (1+l+\Delta )
\end{align}
where $\Re(l+\Delta)>-1$. In the above integral, a small deformation is introduced $k\eta \rightarrow (1 \mp i\xi)\,k\eta$, to account for the time- and anti-time-ordering contours. This deformation arises naturally within the Schwinger--Keldysh formalism, which prescribes the correct contour in the path-integral representation of expectation values (see Appendix~\ref{appendixkeldysh} for details). The same integral structure appears in the de~Sitter scenario. For the non-local terms of the mixed seed function, we have
\begin{align}
\lim_{s\to 0}\mathcal{\hat{I}}^{l_1,l_2} 
=  \frac{H^2 s^{-2}}{\gamma^{3/2}\pi}\bigg\{ 
&
\bigg( \frac{\gamma s^2}{4}\bigg)^{\Delta_+}
\Gamma(-i\mu/2)^2 
\bigg[J^{l_1,\Delta_+}_{+}(k_{12})+J^{l_1,\Delta_+}_{-}(k_{12})\bigg] \\
&\times \bigg[J^{l_2,\Delta_{+}}_{+}(k_{34})
+J^{l_2,\Delta_{+}}_{-}(k_{34})\bigg] + 
\big(\Delta_{+}\to \Delta_{-}, \mu \to - \mu \big)
\bigg\} \ . 
\nonumber
\end{align}
We can recast the product of $J^{l_1,\Delta_\pm}_{\pm}$ as
\begin{align}
    &\big[J^{l_1,\Delta_+}_{+}(k_{12})+J^{l_1,\Delta_+}_{-}(k_{12})\big] \big[J^{l_2,\Delta_+}_{+}(k_{34})+J^{l_2,\Delta_+}_{-}(k_{34})\big] \nonumber\\
    &= -\bigg\{[-i(k_{12})]^{-1-l_1+\Delta_+}\Gamma(1+l_1+\Delta_+)-[i(k_{12})]^{-1-l_1+\Delta_+}\Gamma(1+l_1+\Delta_+)\bigg\}\nonumber\\
    & \quad \times \bigg[ k_{12}\to k_{34}, \ l_1 \to l_2 \bigg].
\end{align}

These integrals give rise to nonlocal terms that have an oscillatory behavior due to the mass of the fields in the bulk, which characterizes particle production.~There are three important cases for the mixed seed: 
\begin{itemize}
    \item the case where \(l_1 = l_2 = 0\) reproduces the amplitude of the trispectrum originating from the interaction $(\partial_{\eta}\varphi)^2\sigma$;
    \item the case where \(l_1 = l_2 = -2\) corresponds to next-to-leading-order corrections to the power spectrum \cite{Aoki:2023wdc} and also includes amplitudes involving external conformal de Sitter particles with mass $m^2 = 2H^2$;
    \item The last case where \(l_1 = 0, l_2 = -2\) is particularly relevant for inflationary phenomenology, allowing connections to cosmological observables.
\end{itemize}
The last two cases are especially useful for comparing with standard de Sitter scenarios previously considered in the literature \cite{Arkani-Hamed:2018kmz, Baumann:2019oyu, Arkani-Hamed:2015bza}. Let us take the case $l_1=l_2=-2$.~The mixed seed in the collapsed limit is the following.
\begin{align}
\lim_{s\to 0}\mathcal{\hat{I}}_{\text{mix}}^{-2,-2} 
= \frac{H^2 \sqrt{u v} }{4\pi}\bigg\{ \bigg( \frac{u v\gamma}{4}\bigg)^{i\mu}&\big[1+i\sinh{(\pi \mu)} \big]  \Gamma(-i\mu/2)^2 \Gamma(1/2 + i\mu)^2 
+\text{c.c.}\bigg\} \quad .
\label{collapsedmixed}
\end{align}
In the above expression, we have used the following definitions of the kinematical variables,
\begin{equation}\label{kinvar}
u \equiv \frac{s}{k_{1}+k_2}\quad , \quad v \equiv \frac{s}{k_{3}+k_4} 
\quad  . 
\end{equation}

The de Sitter seed was first computed in \cite{Arkani-Hamed:2015bza} and re-obtained in several other works \cite{Arkani-Hamed:2018kmz,Baumann:2019oyu,Aoki:2023wdc,Pimentel:2022fsc},
\begin{align}
\lim_{u,v\to 0}\mathcal{\hat{I}}_{dS}^{-2,-2}
= \frac{H^2 \sqrt{u v} }{2\pi}\bigg\{ \bigg( \frac{u v}{4}\bigg)^{i\mu}&\big[1+i\sinh{(\pi \mu)} \big]  \Gamma(-i\mu)^2 \Gamma(1/2 + i\mu)^2 
+\text{c.c.}\bigg\} \quad ,
\label{desitterseedl2}
\end{align}

The relevant kinematical variables are functions of the momentum cross-ratios that characterize the theory. 
In both the mixed and de~Sitter cases, these cross-ratios are constrained by conformal symmetry and therefore coincide, reinforcing the idea that boundary symmetries essentially determine the structure of the solutions. Despite the modification introduced by the term $k^4$ in the ghost equations of motion, the fields appear to share the same behavior at the boundary in both cases.

\begin{figure}[t]
\centering
\hfill
\includegraphics[width=6cm]{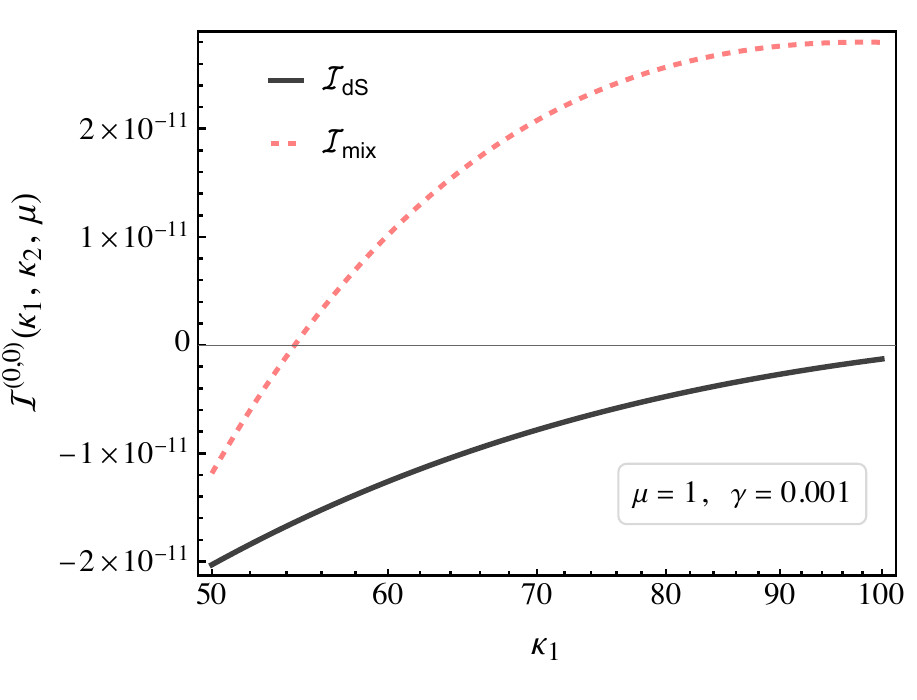}
\hfill
\includegraphics[width=6cm]{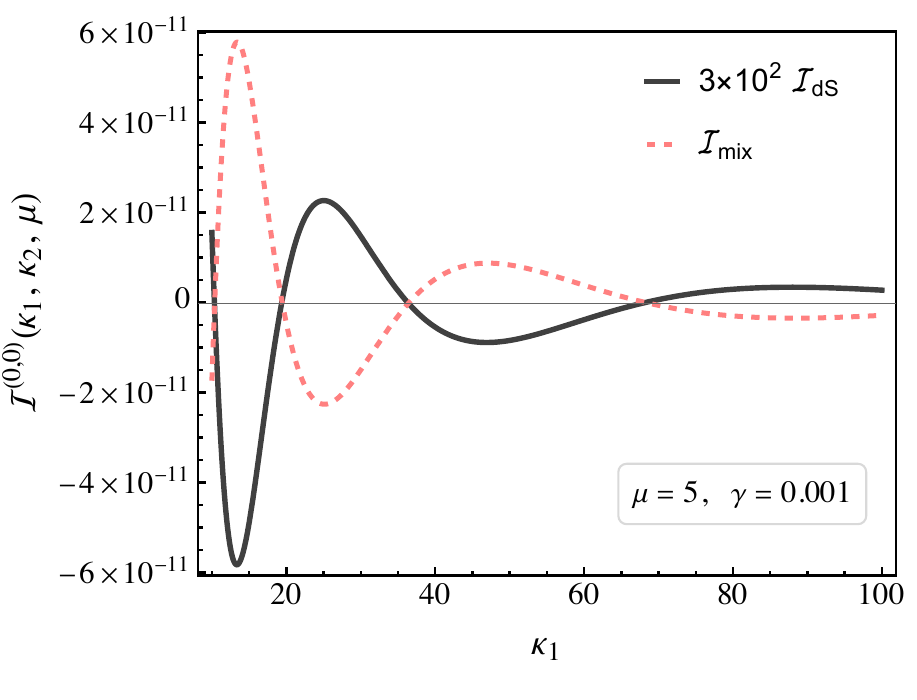} 

\caption{Oscillatory (non-local) contribution to the trispectrum in the collapsed limit, where 
\( \kappa_{1} = u^{-1} \) and \( \kappa_{2} = v^{-1} \). 
In the left panel, the black curve shows the full de Sitter result 
\( \hat{\mathcal{I}}^{0,0}_{\mathrm{dS}}(\kappa_{1},\kappa_{2}) \) for \( \mu = 1 \), 
while the red dashed curve shows the mixed seed 
\( \hat{\mathcal{I}}^{0,0}_{\mathrm{mix}}(\kappa_{1},\kappa_{2}) \) for the same value of \( \mu \). 
For \( \mu = 1 \), both contributions have the same order of magnitude. 
As shown in the right panel, when the mass is increased to \( \mu = 5 \), 
the de Sitter seed must be multiplied by a factor of \( 10^{2} \) in order to be comparable to the mixed result. 
In this analysis we have fixed \( v = 0.005 \).}
\label{trispectro}
\end{figure}

\begin{figure}[t]
\centering
\includegraphics[width=7cm]{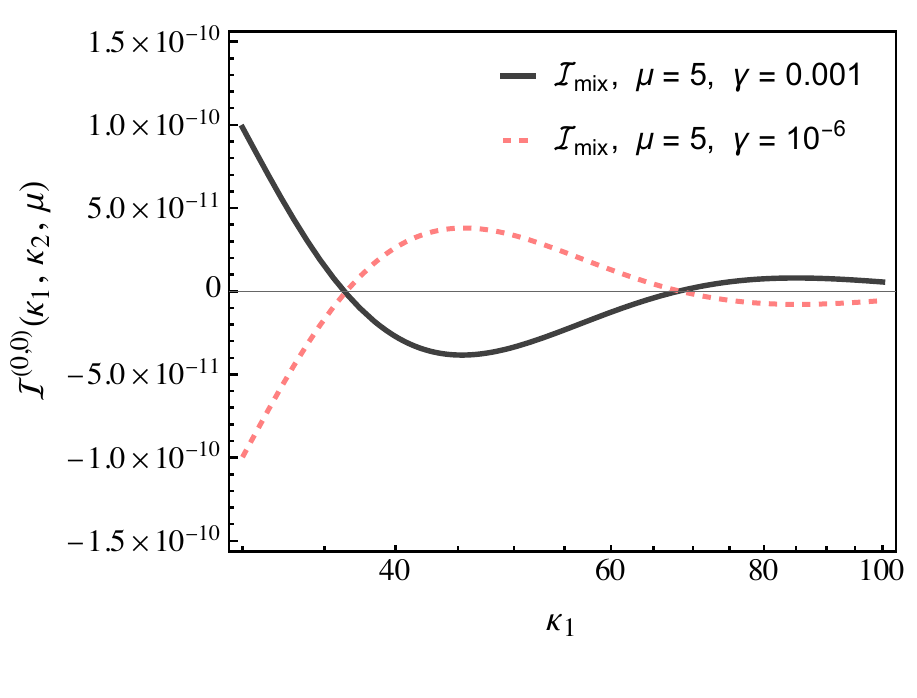}

\caption{Mixed seed function 
\( \hat{\mathcal{I}}^{0,0}_{\mathrm{mix}}(\kappa_{1},\kappa_{2}) \) 
for \( \mu = 5 \). The black solid curve corresponds to $\gamma = 0.001$, and the red dashed curve corresponds to \( \gamma = 1 \cdot 10^{-6} \), where we also have set $v=0.005$.~The amplitude does not change; this shows that the effect of $\gamma$ is to shift the phase, acting as an effective sound speed.}
\label{trispectro2}
\end{figure}

If we consider the ansatz $\sigma(\eta, k) \sim \mathcal{O}(k)\,\eta^{\Delta}$ and take the limit $\eta \to 0$ in Eq.~\eqref{ghostdynamicsbulk}, the resulting anomalous operator takes the same form as in Eq.~\eqref{anomalousdimen}. This suggests that within the ghost seed, conformal symmetry is effectively shifted to different anomalous dimensions. What corresponds to the conformal weights $\Delta = 2, 1$ in the de~Sitter case now corresponds to $\Delta = \tfrac{5}{2}, \tfrac{1}{2}$ in the ghost counterpart.~It is interesting to see if in the collapsed limit, where the signature of new particles appears, there is a sharp difference between the two scenarios. The trispectrum is given by,
\begin{equation}
    \mathcal{T} = \frac{1}{(2 \pi)^6 P^3_{\zeta}} \frac{H^4}{\dot{\phi}^4}\frac{(k_1k_2k_3k_4)^4}{(k_1+k_2+k_3+k_4)^3}\left<\varphi_{\bf k_1}\varphi_{\bf k_2}\varphi_{\bf k_3} \varphi_{\bf k_4}
\right>^{\prime}_\sigma.
\end{equation}
As mentioned earlier, this trispectrum corresponds to the case of massless fields at the boundary, that is, when $l_{1} = l_{2} = 0$. The four-point function is proportional to the seed function (\ref{Seedgeral}), which is the object that carries the oscillatory behavior and captures the difference between the two scenarios. For small values of $\mu$ the difference in amplitude is mild, but as we increase the mass of the exchanged field the signal grows significantly.

This behavior is shown in Fig.~\ref{trispectro}, where we define the variables \( \kappa_{1} = u^{-1} \) and \( \kappa_{2} = v^{-1} \). We compare the de Sitter case (black curve) with the mixed case (blue curve) for \( \mu = 1 \). In this regime, both amplitudes have similar magnitudes, of order \( \mathcal{O}(10^{-11}) \). However, as the mass increases, a significant difference emerges. For example, at \( \mu = 5 \), the difference is \( \mathcal{O}(10^{3}) \). The trispectrum with ghost exchange is therefore amplified and this amplification becomes even stronger for higher values of the exchanged mass.~Including the analytic/local terms does not qualitatively change this enhancement.

The parameter \( \gamma \) does not significantly modify the overall amplitude of the trispectrum. In fact, it can be interpreted as an effective speed of sound, which allows the solutions to shift phase, as illustrated in Fig.~\ref{trispectro2}. Physical arguments (see discussion after \eqref{L2sigmaaction}) show that $\gamma\ll1$, and hence in this scenario the effective speed of sound is very small. This effect is not entirely transparent at the level of the trispectrum; however, it becomes substantially more evident upon examining the bispectrum, as we shall discuss in the following section.

%%%%%%%%%%%%%%%%%%%%%%%%%%%%%%%%%%%%%%%
%%%%%%%%%%%%%%%%%%%%%%%%%%%%%%%%%%%%%%%
%%%%%%%%%%%%%%%%%%%%%%%%%%%%%%%%%%%%%%%
%%%%%%%%%%%%%%%%%%%%%%%%%%%%%%%%%%%%%%%
\section{Bispectrum and the Collider Signal}

To explore the phenomenology of the model, we now focus on the impact of the ghost spectator field on the inflationary bispectrum. The phenomenological effects of the exchanged particle are manifest in the squeezed limit of the bispectrum when the momentum of the internal propagator goes to zero. We consider the following possible interactions:
\begin{align}
&\delta \mathcal{L}_2 \supset \partial_{\eta}\varphi \, \sigma(\eta)
\quad , \qquad 
\delta \mathcal{L}_3 \supset (\partial_{\eta}\varphi)^2 \sigma(\eta)
\quad ,
\end{align}
for (external) massless coupled fields. The three-point function reads
\begin{equation}\label{integregr}
    \langle  \varphi_{\bf k_1} \varphi_{\bf k_2}\varphi_{\bf k_3} \rangle^{\prime}_{\sigma} = \frac{H}{4 k_1 k_2 k_3} \sum_{ab = \pm} \int d\eta_1 \frac{d\eta_2}{(-\eta_2)^2} e^{i a k_{12} \eta_1 + ib k_3 \eta_2} G_{ab}(s,\eta_1,\eta_2) + 2\ \text{perm,}
\end{equation}
In the squeezed limit, the small-$s$ expansion isolates only the soft exchange channel, 
which carries the non-analytic oscillatory cosmological-collider signal, whereas the 
remaining channels contribute to the analytic hard background.\footnote{More explicitly, 
the full exchange bispectrum can be decomposed as 
$\langle \phi_{\mathbf{k}_1}\phi_{\mathbf{k}_2}\phi_{\mathbf{k}_3}\rangle'_\sigma
= C_{12|3}+C_{23|1}+C_{31|2}$.
In the squeezed configuration $k_3\ll k_1\sim k_2$, one has 
$s_{12}=k_3$, $s_{23}=k_1$, and $s_{31}=k_2$. Therefore only $C_{12|3}$ probes the 
soft limit $s\to0$, while $C_{23|1}$ and $C_{31|2}$ remain hard.}
We can interpret \eqref{integregr} as the limit $k_4\rightarrow 0$ of the four-point function. In other words, the bispectrum can be obtained by taking the soft limit of the seed integral~\cite{Aoki:2023wdc,Qin:2023ejc,Yin:2023jlv}
\begin{align}
& \langle  \varphi_{\bf k_1} \varphi_{\bf k_2}\varphi_{\bf k_3} \rangle^{\prime}_{\sigma}  = -\frac{H}{4 k_1 k_2 k_3^4} \lim_{k_4\to 0} \sum_{a,b=\pm}\mathcal{\hat{I}}^{0,-2}_{ab} + 2\ \text{perm.}\notag
\end{align}
The above expression is quite general, meaning that a slow-roll bispectrum can be obtained from a soft limit of the four-point function.~Recall that the curvature perturbation is obtained from scalar field perturbations by the relation $\zeta = \varphi / (\sqrt{2\epsilon}\, M_{\mathrm{Pl}})$.~The corresponding non-analytic contribution to the bispectrum $B(k_1,k_2,k_3)_{\sigma} \equiv \langle \zeta_{\bf k_1}\zeta_{\bf k_2}\zeta_{\bf k_3} \rangle^{\prime}$ can be parameterized by a 
reduced shape function $S$, defined through
\begin{equation}
    \langle \zeta_{\bf k_1}\zeta_{\bf k_2}\zeta_{\bf k_3} \rangle^{\prime} =(2\pi)^4 \frac{P^{2}_{\zeta}}{( k_1 k_2 k_3)^2} S
\end{equation}
where $P_{\zeta} = H^2/8\pi^2 \epsilon M_{pl}^2$ is the power spectrum.~Here we only show the result of S in the squeezed
limit $k_3 \ll k_2 \sim k_1$.~The result is
\begin{equation}
S_{\text{mix}} = 
\frac{1}{(2\pi)^4} \cdot \frac{1}{P^2_{\zeta}}\cdot (2 \epsilon M_{pl}^2)^{-\frac{3}{2}} \cdot \frac{H^3}{16 \pi} \, F_{\text{mix}}(\kappa,\mu) 
\end{equation}
with 
\begin{equation}
F_{\text{mix}}(\kappa,\mu) \equiv 
\frac{e^{ -\pi \mu }}{2\kappa^{1/2}} \bigg\{\bigg(\frac{\gamma}{\kappa}\bigg)^{i \mu}\frac{i(i-e^{\pi\mu})^2}{2^{1 + 2i \mu}}\Gamma \left(i \mu +\frac{1}{2}\right) \Gamma \left(i \mu +\frac{5}{2}\right) \Gamma \left(-\frac{i \mu}{2}\right)^2 + \textbf{c.c} \bigg\} \quad ,
\end{equation}
where $\kappa \equiv k_1/k_3 \approx1/2u$. For comparison, we can calculate the de Sitter shape function for the same interactions.~In this case, we find the following.
\begin{equation}
    S_{\text{dS}} = \frac{1}{(2\pi)^4} \cdot \frac{1}{P^2_{\zeta}}\cdot (2 \epsilon M_{pl}^2)^{-\frac{3}{2}} \cdot \frac{H^3}{16 \pi} \, F_{\text{dS}}(\kappa,\mu) \quad .
\end{equation}
with 
\begin{equation}
    F_{\text{dS}}(\kappa,\mu) \equiv \frac{e^{ -\pi \mu }}{\kappa^{1/2}} \bigg\{\bigg(\frac{1}{\kappa}\bigg)^{i \mu}\frac{i(i-e^{\pi\mu})^2}{2^{1 + 2i \mu}}\Gamma \left(i \mu +\frac{1}{2}\right) \Gamma \left(i \mu +\frac{5}{2}\right) \Gamma \left(-i \mu \right)^2 + \textbf{c.c} \bigg\} \quad .
\end{equation}

Since the boundary fields are the same in both cases, the prefactors are also the same.~The expressions above describe the non-analytic oscillatory part of the bispectrum in the squeezed limit. If one restores the local analytic terms, the full bispectrum is modified, as expected. Nevertheless, the qualitative enhancement of the ghost signal relative to the de Sitter case remains unchanged.~Therefore, following common practice in the literature, we retain only the non-local contribution, as it provides the cleanest way to isolate the genuine cosmological collider signal.~The deviation from the standard de~Sitter result becomes increasingly pronounced for larger values of $\mu$. The spontaneous breaking of Lorentz symmetry induces a modified dispersion relation that substantially amplifies the collider signal. The enhancement in signal comes from the higher–derivative momentum dependence $ k^{4}$,
\begin{equation}
 \omega_{\text{dS}} \sim k, \quad  \omega_{\text{ghost}} \sim k^2. 
\end{equation}
This fact has two main consequences: the Ghost mode is not a Bunch-Davies mode in the usual sense and, when integrated in time, becomes less Boltzmann-suppressed. Some proposals in the literature introduce modifications to the mass of the theory to generate an effective chemical potential that compensates for the Boltzmann suppression and amplifies the signal. Interestingly, the massive ghost field naturally exhibits this characteristic.~We can see this qualitatively as follows: 
\begin{align}
\mathcal{I}_{dS} &\sim O(1) \, \Gamma(-i \mu )^2 \Gamma(i \mu + 5/2 )^2  \approx   \, \frac{2 i \pi  }{\mu } e^{-\mu  (2 i \log (\mu )-2 i+\pi )}\\
\mathcal{I}_{\text{ghost}} 
&\sim O(1) \, \Gamma(-i \mu /2 )^2 \Gamma(i \mu + 5/2 )^2  \approx   \, \frac{4 i \pi  }{\mu }e^{-\frac{1}{2} \mu  \left(2 i \log \left(\frac{\mu }{2}\right)-2 i+\pi \right)} .
\end{align}

\begin{figure}[t]
\hfill
{\includegraphics[width=7cm]{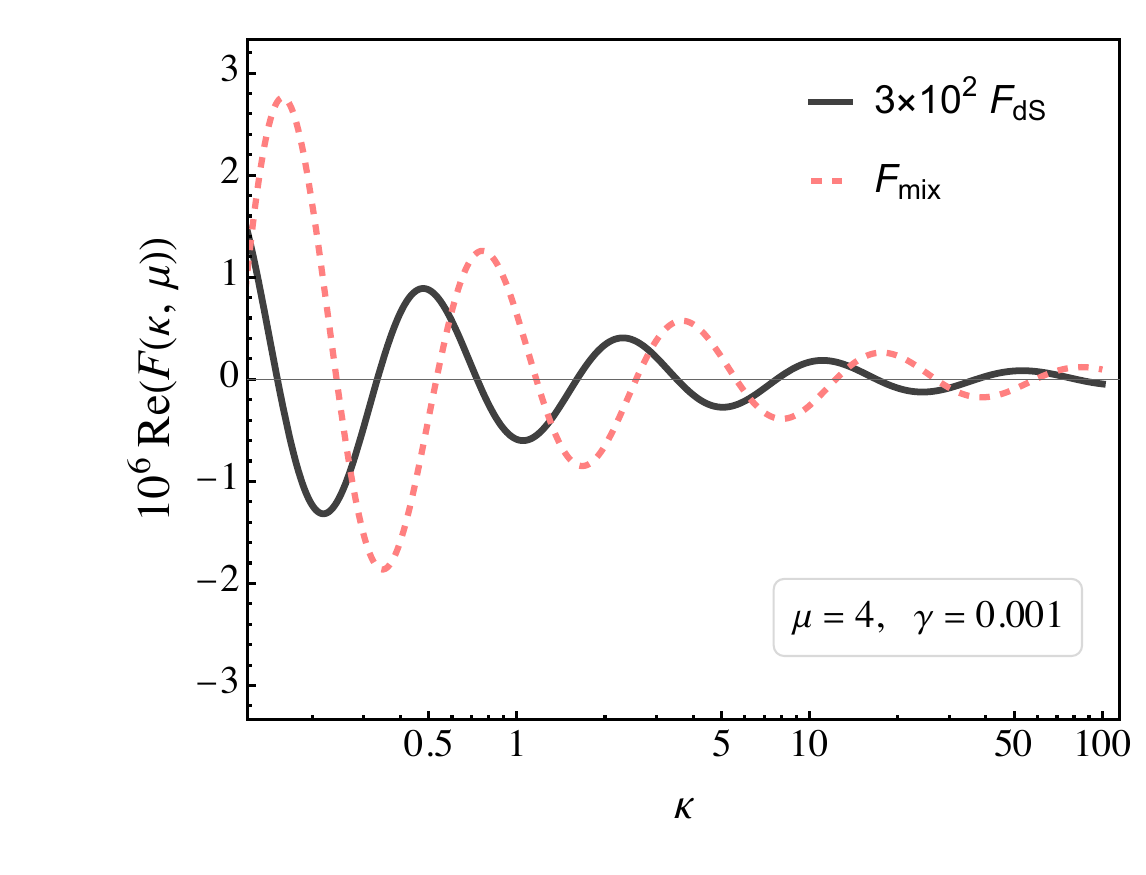}}
\hfill
{\includegraphics[width=7cm]{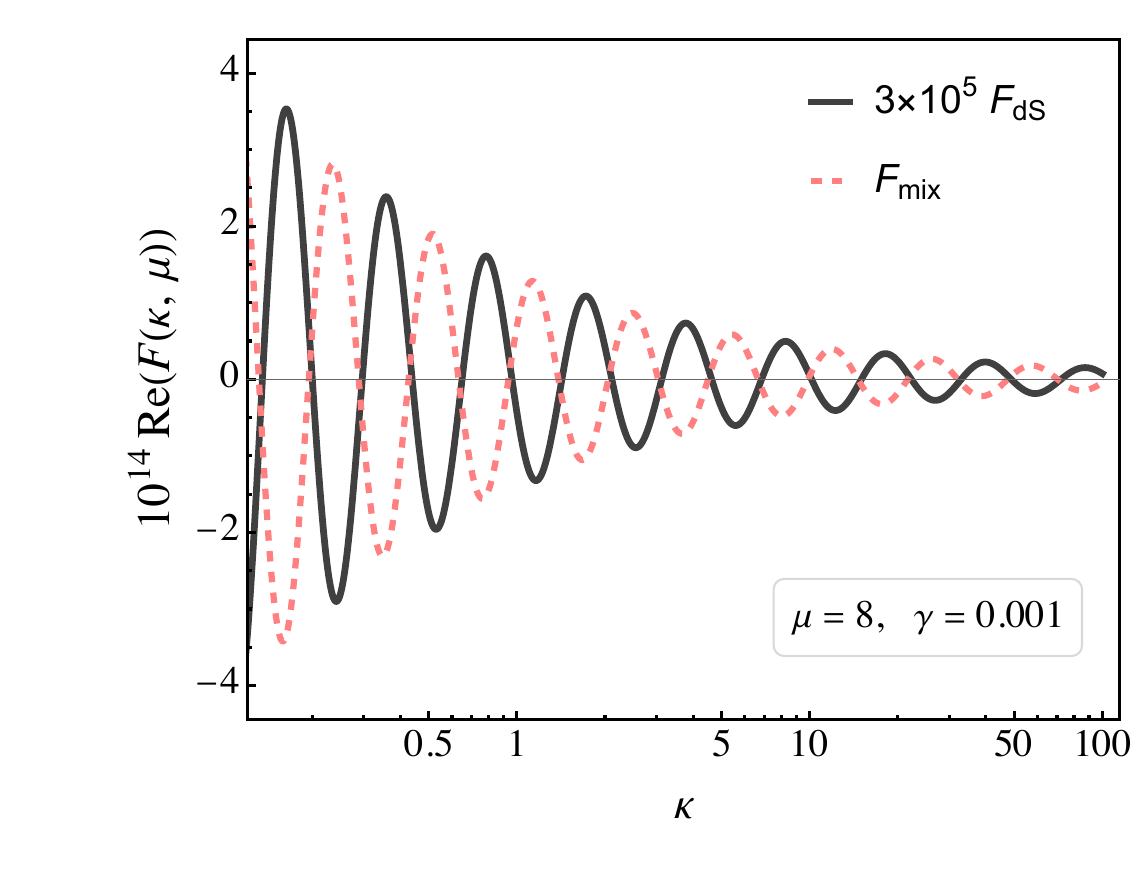}}
\hfill
\caption{This plot shows the role of the mass of the fields, as the mass increases the discrepancy between the amplitudes becomes even bigger.~The oscillatory contribution to the bispectrum.~On the left the comparison between the full dS case $F_{\text{dS}}$ and the mixed $F_{\text{mix}}$ for the exchange of a ghost in the bulk, for $\mu = 4$.~In the right we have considered $\mu = 8$ and for both curves we set $\gamma = 0.001$.}
\label{bISPECATEMPT}
\end{figure}

As the mass of the exchanged particle increases, the magnitude of the bispectrum in the ghost case can differ by up to \( \mathcal{O}(10^{5}) \) (Fig.~\ref{bISPECATEMPT}) from that in the de~Sitter case.~This result is particularly interesting in light of recent developments in the cosmological collider framework involving non-Bunch-Davies initial states~\cite{Yin:2023jlv}, as well as works that introduce a chemical potential to counterbalance the Boltzmann suppression~\cite{Bodas_2021}. In both contexts, significant enhancements in the bispectrum amplitude have been reported.

\begin{figure}[t]
\hfill
{\includegraphics[width=7cm]{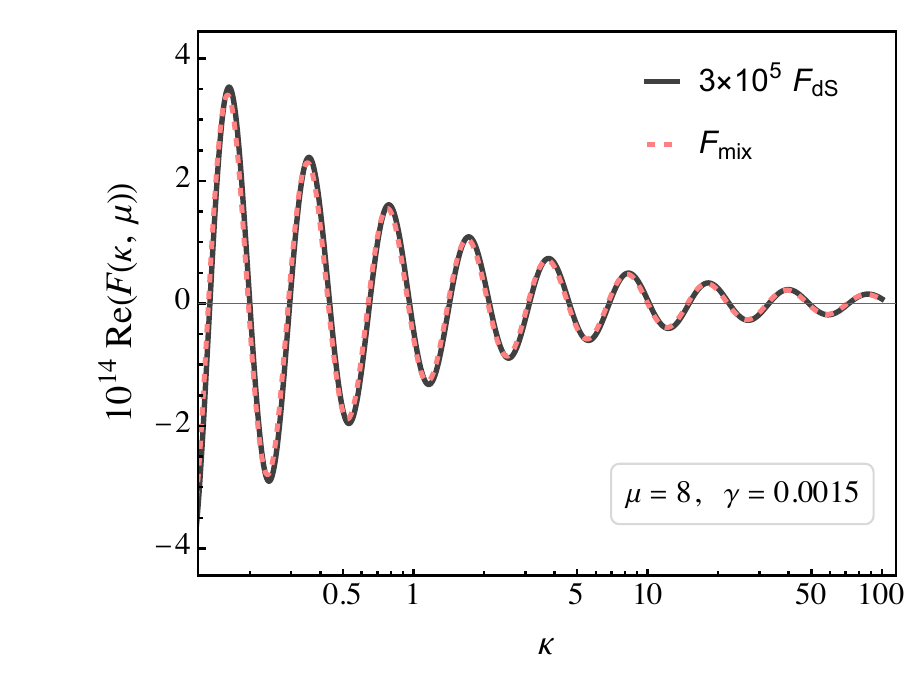}}
\hfill
{\includegraphics[width=7cm]{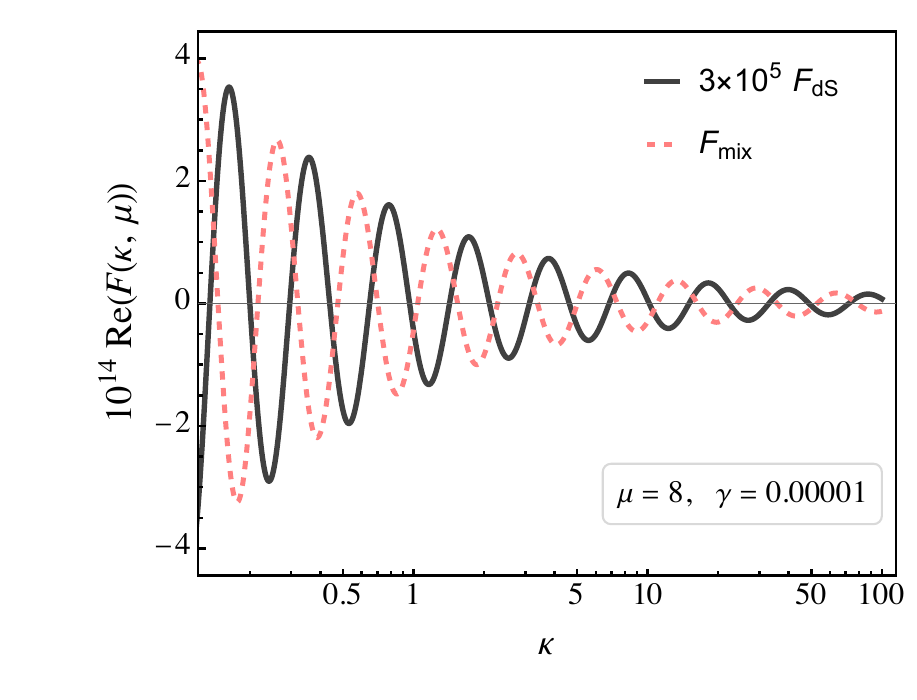}}
\hfill
\caption{This plot shows the role of the gamma $\gamma$ in fields, as gamma changes the signal can be in phase or out of phase, reproducing an "effective" speed of sound.~The oscillatory contribution to the bispectrum.~On the left the comparison between the oscillatory dS case $F_{\text{dS}}$ and the mixed $F_{\text{mix}}$ for the exchange of a ghost in the bulk, for $\mu = 8$ and $\gamma = 15 \cdot 10^{-4}$.~In the right we have considered $\gamma = 1 \cdot 10^{-5}$.}
\label{bISPECATEMPT22}
\end{figure}

It is also important to note that for conformally coupled cases, such as the de Sitter scenario with $m^{2} = 2H^{2}$ and the corresponding ghost case with $m^{2} = \tfrac{5}{4}H^{2}$, the amplification effect becomes even stronger.~The energy scale of the ghost introduces a new parameter $\gamma = H/M$. As mentioned above, $\gamma$ can be effectively interpreted as a small speed of sound, which does not change the overall amplification of the signal.~A small change in this parameter can even place the signal in phase or out of phase with the standard de Sitter case, as shown in Fig.~\ref{bISPECATEMPT22}.

The similarity between the two cases shown in Fig.~\ref{bISPECATEMPT22} holds for any mass of the de Sitter field, that is, for any parameter $\nu=i\mu$ there is a value of $\gamma$ that allows us to map the two signals. We can perfectly map them apart from a normalization constant. However, the parameter $\gamma$ is related to the energy scale of the ghost condensate, so we can adjust this parameter as long as it remains small $\gamma\ll 1$.

\section{Remarks on Bootstrap}

In the preceding section, we analyze the cosmological collider signal associated with the ghost field. To obtain the complete shape function, it is necessary to employ bootstrap techniques \cite{Arkani-Hamed:2018kmz, Baumann:2019oyu, Pimentel:2022fsc, DuasoPueyo:2023kyh}.~This requires solving the bootstrap equations specific to the ghost collider, which have been rigorously derived in the Appendix~\ref{appendixbootsequationsghost}. These equations are given by,
\begin{align}
& \mathcal{G}_{u}^{l_1} \mathcal{I}_{ \pm \mp}^{l_1 l_2}\left(u, v\right)=0 \label{homogeneadiffghost},\\
& \mathcal{G}_{u}^{l_1} \mathcal{I}_{ \pm \pm}^{l_1 l_2}\left(u, v\right) =H^2e^{\mp \mathrm{i} (l_1+l_2) \pi / 2} \Gamma\left(5+l_1+l_2\right)\left(\frac{u v}{u+v}\right)^{5+l_1+l_2} .
\end{align}
where the operator $\mathcal{G}^{l}_{u}$ is given by,
\begin{align}
\mathcal{G}^{l}_{u}\equiv  \mu ^2+\frac{1}{4} (2 l+5)^2
&-2 \left[(l+2)u-12 \gamma ^2 u^5\right]\partial_{u} 
\notag\\
&   + (u^2 + 36u^6\gamma^2)\partial_{u}^2 + u^7 \gamma^2(12 \partial_{u}^3 + u\partial_{u}^4).\label{bootdiffope}
\end{align}
In contrast to the standard de~Sitter bootstrap equations, the modified dispersion relations derived here exhibit an explicit dependence on higher-derivative terms. This distinctive feature of the formalism, namely the role of higher-derivative contributions within bootstrap frameworks, has not yet been systematically investigated in the literature.~There are several important features of this differential equation. Since the conformal symmetry is broken in the bulk, one might expect changes in the boundary behavior. However, as discussed earlier, the boundary limit \( \eta \to 0 \) still reproduces the same structure as in the full de Sitter case. This is reflected in the constraint equation,
\begin{equation}
   \big( \mathcal{G}_{u}^{l_1} - \mathcal{G}_{v}^{l_2} \big)
   \left( \frac{u v}{u+v} \right)^{5 + l_1+l_2} = 0 .
\end{equation}

The above expression remains unchanged when $l_{1} = l_{2}$, in agreement with the standard de Sitter result. This indicates that, although Lorentz invariance is explicitly broken, a nontrivial subgroup of the conformal symmetry remains preserved and continues to constrain the structure of the differential equation. In this sense, conformal symmetry is still partially encoded in the solution, even though its full de Sitter realization is lost. The main difficulty associated with Eq.~\eqref{bootdiffope} lies in its singularity structure. In the usual de Sitter case, the boundary conditions are fixed by the singular limits $u \to \pm 1$, which are directly tied to factorization and collinear limits of the correlators. In the present setup, where Lorentz symmetry is absent, it is not immediately clear whether these singularities persist or how they should be interpreted. The differential equation~\eqref{homogeneadiffghost} is a generalized hypergeometric equation, whose general solution is given by a linear combination of four independent hypergeometric functions. Determining which combinations are physically admissible therefore requires a careful analysis of the modified singularity structure. What complicates the analysis is that the correct boundary conditions are not evident and determining the proper normalization of the solutions is nontrivial. A more detailed study of these solutions and their physical implications is left for future work.

The structure of these equations raises an important question regarding the appropriate boundary conditions of the model. The only manifest singularity appears in the limit \( u \to 0 \), which corresponds to the particle production signal in the collapsed limit of the trispectrum. In contrast, the two additional singular points present in the de Sitter case seem to be absent here, suggesting that the singularity structure of the ghost correlator may differ in a fundamental way.~One of these is associated with a well-defined Bunch-Davies vacuum amplitude. This observation suggests the possibility of a non-trivial connection between the flat-space limit of the ghost modes and a non-Bunch-Davies solution.

In the broader context, models of ghost inflation exhibit a remarkable feature related to the pursuit of ultraviolet completeness, often regarded as a step toward a unified theoretical framework. Although the present analysis does not yield a complete analytical solution, the issues concerning the proper definition of the vacuum and the effective field-theoretic description may have other interesting structures. Addressing these aspects could ultimately provide insight into the correct formulation of the bootstrap equations for this scenario.

\clearpage

\section{Conclusion}

In this work, we developed a new perspective on the Cosmological Collider program by considering particles with modified dispersion relation during inflation. Although this particle cannot remain as an independent degree of freedom outside the horizon, it can still leave an imprint in the non-Gaussian sector of primordial correlators. The effect arises from the exchange of a massive ghost condensate excitation in the inflationary bulk. 

Even in the presence of explicit Lorentz violation, the collapsed limit of the correlators preserves the characteristic oscillatory pattern known from the standard de Sitter scenario. Moreover, the absence of a quadratic term in the dispersion relation $\omega^2 \propto k^4$ generates Boltzmann enhancement, thus providing a scenario having both dS and boostless characteristics. The parameter $\gamma = H/M$, associated with the underlying ghost condensate, plays an important role in this scenario%enhancement
.~It controls the relative phase of the oscillatory contribution, allowing the ghost mediated signal to appear either in phase or out of phase with respect to the standard de Sitter result. Despite these bulk modifications, the correlators remain almost scale-invariant. The bootstrap equations are new, and it would be interesting to elucidate their symmetry structure and find their closed-form solutions.

These findings have direct implications for cosmological observations. The enhanced amplitude of the non-Gaussian signal effectively shifts the energy scale at which heavy-particle signatures become observable. This creates a promising opportunity for the detection of such signals in upcoming surveys of large-scale structure and in future 21-cm observations. The amplification mechanism presented here may, therefore, bring high-mass collider signatures into the range of realistic observational prospects.

The scenario developed in this work demonstrates that ghost-condensate excitations provide a compelling mechanism for amplifying cosmological collider observables. The reduced suppression of heavy-particle effects, combined with the preservation of the characteristic oscillatory structure at the boundary, establishes ghost-mediated non-Gaussianity as a realistic and potentially detectable probe of heavy new physics in the primordial universe.

\enlargethispage{2\baselineskip}
\acknowledgments 
MCF and FTF acknowledge the financial support of the National Scientific and Technological Research Council (CNPq, Brazil) and the Brazilian Federal Agency for Support and Evaluation of Graduate Education (CAPES, Brazil). MCF thanks the Scuola Normale Superiore for hospitality during part of this project. A first version of this project was presented at ``QCD meets Gravity 2025" by MCF, and we thank the participants for questions and feedback.

GLP is supported by Scuola Normale, by INFN (IS GSS-Pi), and by the ERC (NOTIMEFORCOSMO, 101126304). Views
and opinions expressed are, however, those of the author(s) only and do not necessarily reflect
those of the European Union or the European Research Council Executive Agency. Neither the
European Union nor the granting authority can be held responsible for them. GLP is further supported by the Italian Ministry of Universities and Research (MUR) under contract 20223ANFHR
(PRIN2022).

%%%%%%%%%%%%%%%%%%%%
%%%%%%%%%%%%%%%%%%%%
%%%%%%%%%%%%%%%%%%%%

\pagebreak
\appendix
\section{Mode Functions}
\label{appendix::besselfunctions}
The equations of motion can be written in terms of the generalized Bessel's equation \cite{bowman1958introduction},
\begin{equation}
    \big[ x^2 \partial_{x}^2 + (2p+1)x\partial_{x}+\big( a^2 x^{2r}+\beta^2 \big)\big]y(x) = 0.
\end{equation}
Its solution is given by,
\begin{equation}
    y(x) = x^{-p}\bigg[ c_1 J_{q/r}\bigg(\frac{a}{r} \ x^r\bigg)+c_2 Y_{q/r}\bigg(\frac{a}{r} \  x^r\bigg) \bigg] 
\end{equation}
where $q \equiv \sqrt{p^2 - \beta^2}$ and $c_1$ and $c_2$ are constants to be determined. Rewrite the Bessel functions in terms of the Hankel functions (or Bessel functions of the third kind) defined as follows:
\begin{equation}
  \mathcal{H}^{(1)}_\nu(z) = J_\nu(z) + iY_\nu(z), \quad 
\mathcal{H}^{(2)}_\nu(z) = J_\nu(z) - iY_\nu(z),  
\end{equation}
we can directly determine the constants \( c_1 \) and \( c_2 \) by imposing the condition that the modes should at early-times behave like plane waves in Minkowski space (scaling as \( e^{iz} \)),
\begin{equation}
    \lim_{\eta \to - \infty} u_k  = \frac{e^{-ik\eta}}{\sqrt{2k}}.
\end{equation}
By using this boundary condition, we obtain the relation \( c_2 = i c_1 \). Here, \(|c_1| \) is obtained by the normalization condition:
\begin{equation}
    u_k\partial_{\eta}u_k^* - u_k^*\partial_{\eta}u_k = -i .
\end{equation}
The choices $p=-1/2$, $r=1$ and $p=-1/2$, $r=2$ reproduce the de Sitter and ghost modes.~The asymptotic limits of the Hankel functions are the following,
\begin{align}
    &\lim_{z \to \infty}\mathcal{H}^{(1)}_{\nu}(z) = \sqrt{\frac{2}{\pi z}}e^{i(z-\frac{1}{2}\nu \pi -\frac{1}{4}\pi)}, \quad \lim_{z \to \infty}\mathcal{H}^{(2)}_{\nu}(z) = \sqrt{\frac{2}{\pi z}}e^{-i(z-\frac{1}{2}\nu \pi -\frac{1}{4}\pi)},
\end{align}
and also,
\begin{equation}
    \lim_{z \to 0}\mathcal{H}^{(1)}_{\nu}(z) = \frac{2^{-\nu } (1+i \cot (\pi  \nu )) z^{\nu }}{\Gamma (1+\nu)}-\frac{i 2^{\nu } \Gamma (\nu ) z^{-\nu }}{\pi }\nonumber
\end{equation}
where using that,
\begin{equation}
    \Gamma(1-\nu)\Gamma(\nu) = \frac{\pi}{\sin(\pi \nu)},\nonumber
\end{equation}
we have,
\begin{equation}
    \lim_{z \to 0}\mathcal{H}^{(1)}_{\nu}(z) =-\frac{i}{\pi}\bigg[  e^{-i \pi  \nu } \Gamma (-\nu ) \bigg( \frac{z}{2}\bigg)^{\nu}+ \Gamma (\nu )\bigg( \frac{z}{2}\bigg)^{-\nu}\bigg].
\end{equation}
The above holds for the standard de Sitter case, Eq.~(\ref{dsmodeHankel}).~For the ghost particle the procedure is analogous, the only difference will lie in the normalization of the constants.~The Bunch-Davies mode function has the following form in
the late-time expansion \cite{Yin:2023jlv},
\begin{equation}
    \lim_{k \to 0} u_k (\eta) = -i \sqrt{\frac{2}{\pi k^3}} H e^{i \pi/4}\left[ e^{-\mu \pi/2} \Gamma(i\mu) \left(-\frac{k\eta}{2} \right)^{3/2 - i\mu} + e^{\mu \pi/2} \Gamma(-i\mu) \left(-\frac{k\eta}{2} \right)^{3/2 + i\mu} \right].
\end{equation}
Note that, for massless objects only one of the contribution survives (the second term will go to zero since $k \to 0$). That is, if $i\mu \equiv \nu = 3/2$ we recover the result of the discussion of Section \ref{sec:modes-and-powerspec} for the massless full de Sitter particles.

%%%%%%%%%%%%%%%%%%%%%%%%%%%%%%%%%%%%%%
%%%%%%%%%%%%%%%%%%%%%%%%%%%%%%%%%%%%%%
%%%%%%%%%%%%%%%%%%%%%%%%%%%%%%%%%%%%%%
%%%%%%%%%%%%%%%%%%%%%%%%%%%%%%%%%%%%%%
\section{Schwinger-Keldysh Formalism}
\label{appendixkeldysh}

\begin{figure}[!ht]
    \centering
    \includegraphics[width=0.55\textwidth]{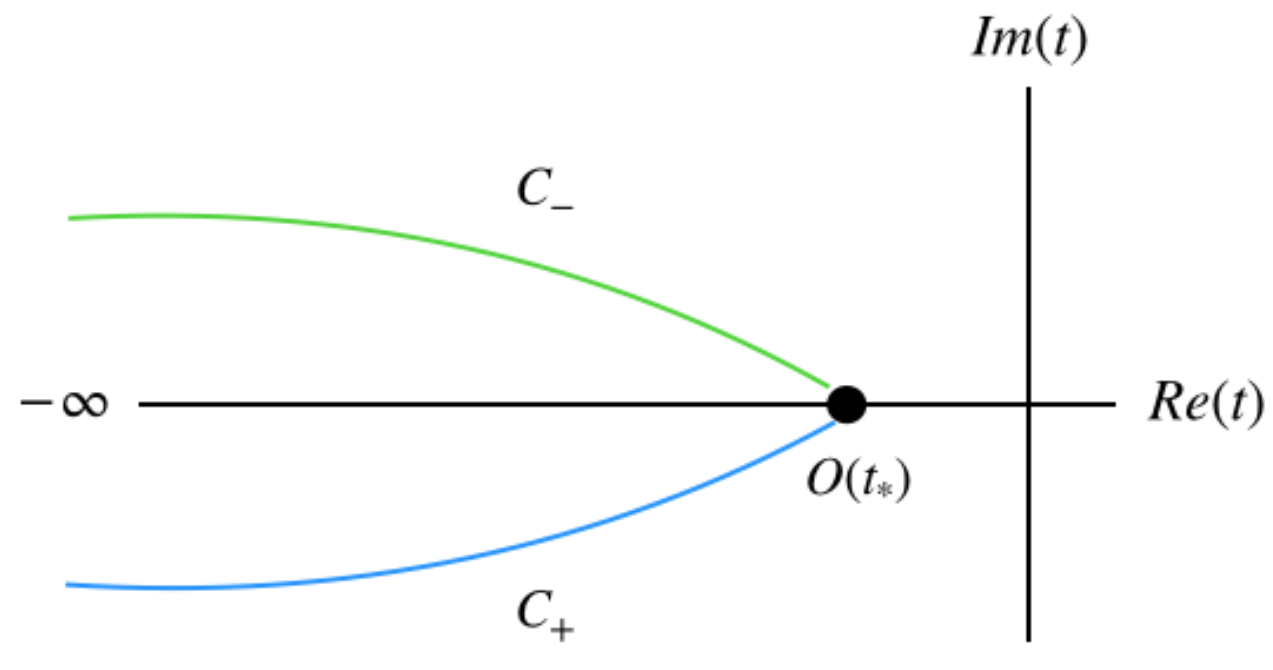}
    \caption{Keldysh contour. The contour $C_{+}:-\infty(1 - i\xi)$ is represented by the green curve; it represents the contour of time-ordered operators. The contour $C_{-}:-\infty(1 + i\xi)$ is represented by the blue curve; it represents the contour of anti-time-ordered operators.}
    \label{keldysh}
\end{figure}

In this formalism, the expectation value is defined as:
\begin{equation}\label{media}
    \langle O(t) \rangle =  \langle \Omega | O(t)| \Omega \rangle ,
\end{equation}
where $t$ represents the cosmic time. The above operator can be a single operator or a product of fields, all evaluated at the same time. The state $|\Omega \rangle$ denotes the vacuum state of the interacting theory. It is preferable not to be confined to a single reference frame where the states vary with time and span the entire space. By decomposing the interaction in the Heisenberg picture, we obtain:
\begin{equation}
    O(t) = U^{\dagger}(t,t_0) O(t_0)U(t,t_0),
\end{equation}
By defining an interaction as $ H_{int}$ we have $ H_0$, from which the operator $O(t_0)$ evolves. This leads us to consider the conditions under which an operator evolves:
\begin{equation}
    O^{I}(t) = U^{\dagger}(t,t_0) O(t_0)U(t,t_0),
\end{equation}
where $t_0$ is the initial time and $U$ is a time-evolution unitary operator,
\begin{equation}
    \frac{dU}{dt}=-iHU, \quad U(t_0,t_0)=1.
\end{equation}
Now, we can re-write as Eq.~(\ref{media}),
\begin{align}
    \langle \Omega |O(t) | \Omega \rangle &= \langle |\Omega U^{\dagger}(t,t_0)O(t) U(t,t_0) | \Omega \rangle\nonumber \\
    &= \langle|\Omega F^{\dagger}(t,t_0) U^{\dagger}_0(t,t_0)O(t) U_0(t,t_0) F(t,t_0)| \Omega \rangle\nonumber\\
    &= \langle \Omega | F^{\dagger}(t,t_0)O^{I}(t)F(t,t_0)| \Omega \rangle
\end{align}
where,
\begin{equation}
    F = T e^{-i\int_{t_0}^{t}dt^{\prime}H^{I}_{int}(t^{\prime})}, \quad F(t_0,t_0) = 1,
\end{equation}
or satisfy unitary time evolution. The operator $T$ denotes time-ordering. Then, we have the in-in master formula
\begin{equation}
    \langle O(t)\rangle = \langle 0  | \hat{T}e^{i\int_{t_0}^{t}dt^{\prime}H^{I}_{int}(t^{\prime})}O(t)Te^{-i\int_{t_0}^{t}dt^{\prime}H^{I}_{int}(t^{\prime})}| 0 \rangle
\end{equation}
where $\hat{T}$ denotes an anti-time-ordering. Let us briefly comment about the importance of the time-ordering operator. Unlike standard quantum field theories, in this in-in approach to cosmology we do not Fourier transform the time variable.~Then the contour integration is in time, not in $E_p$ like standard QFT (See \cite{Peskin:1995ev}). However, the procedure is almost the same. We slightly rotate time to cover the poles $t_{\pm}\equiv (1\mp i\xi)t$, where $\xi>0$, see Fig.~\ref{keldysh}.~This contour is implemented by the time- and anti-time-ordering operators.
These objects will act as
\begin{align}
    T \{ O(t_1)O(t_2)\} = \bigg\{ \begin{array}{cc}
        O(t_1)O(t_2) & t_1 > t_2 \\
        O(t_2)O(t_1) & t_1 < t_2
    \end{array} \bigg\},
\end{align}
or,
\begin{align}
    T \{ O(t_1)O(t_2)\} =&O(t_1)O(t_2)\theta (t_1 - t_2)+O(t_2)O(t_1)\theta (t_2 - t_1),
\end{align}
in order to cover the poles described in Figure \ref{keldysh}. As a consequence, the propagator will be represented by a matrix of propagators. The above discussion also holds for the conformal time $d\eta^2 = a^2(t)dt^2$. To compute the equations of motion for the propagators Eqs. (\ref{diafg}) and~(\ref{nondifr}), we need the expectation value in the path integral representation \cite{Chen:2017ryl}. In the path integral representation, the expectation value is:
\begin{align}
    \langle O(t)  \rangle  = &  \int D\varphi_{+}D\varphi_{-}\varphi_{+}(\eta,x)...\varphi_-(\eta_N,x_N)e^{i\int_{t_0}^{t_f}d\eta d^3\vec{x}\big(\mathcal{L}_{cl}[\varphi_{+}]-\mathcal{L}_{cl}[\varphi_{-}]\big)}\nonumber\\
    &\times\prod_{\vec{x}}\delta\big( \varphi(\eta_f,\vec{x})-\varphi_{-}(\eta_f,\vec{x})\big). 
\end{align}
The above can be computed from the generating functional and by taking functional derivatives.~The procedure can be summarized into a set of diagrammatic rules \cite{Chen:2017ryl}.

\section{Ghost Collider Bootstrap Equations}
\label{appendixbootsequationsghost}
Take the Mukhanov-Sasaki equation (\ref{ghostdynamicsbulk}) for bulk fields. In the flat gauge $v= a \sigma$, the equations of motion will assume the following form,
\begin{equation}
     \bigg[ \eta^2 \partial_{\eta}^2 - 2\eta \partial_{\eta}+\gamma^2s^4\eta^4 + \frac{m^2}{H^2}\bigg]\sigma=0,
\end{equation}
where, $\gamma \equiv \alpha H/M $.~From the above we can obtain the equations of motion for the ghost two-point function,
\begin{align}
    &\bigg[\eta_1^2 \partial_{\eta_1}^2 - 2\eta_1 \partial_{\eta_1}+\gamma^2s^4\eta^4_1 + \frac{m^2}{H^2}\bigg]G_{\pm \mp}(s,\eta_1,\eta_2) = 0\nonumber\\
    &\bigg[ \eta_1^2 \partial_{\eta_1}^2 - 2\eta_1 \partial_{\eta_1}+\gamma^2s^4\eta^4_1 + \frac{m^2}{H^2}\bigg]G_{\pm \pm }(s,\eta_1,\eta_2) =\mp i \eta_1^2 \eta_2^2 \delta (\eta_1 - \eta_2).\label{motionbulk}
\end{align}
Again, it is useful to define the following kinematic variables:
\begin{equation}
    \textbf{s} =  \textbf{k}_1+\textbf{k}_2  =  -( \textbf{k}_3+\textbf{k}_4), \quad u = \frac{s}{k_1+k_2}, \quad v = \frac{s}{k_3+k_4}.
\end{equation}
By doing a variable change $z_1 = k_{12}\eta_1$ and $z_2 =k_{34}\eta_2$, the seed function assumes the following form,
\begin{align}
     &\mathcal{\hat{I}}^{l_1,l_2}(u,v)_{ab} = -ab u^{1+l_1}v^{1+l_2} \int_{-\infty}^{0}dz_1(-z_1)^{l_1} e^{\pm i z_1} \int_{- \infty}^{0}dz_2(-z_2)^{l_2}e^{\pm i z_2}\hat{G}_{\pm \pm} (u z_1, vz_2),
\end{align}
where we are using the dimensionless propagator,
\begin{align}
    \hat{G}(u z, vz^{\prime}) &\equiv s^3G \big(s,z/(k+p),z^{\prime}/(q+r)\big).
\end{align}
Now, instead of performing the integration, we can bootstrap its analytical form.~The equations of motion for the bulk propagators in the kinematic variables are the following ones,
\begin{align}
     &\bigg[ z^2 \partial^2_{z} - 2z \partial_{z}+\gamma^2 u^4 z_1^4+\frac{m^2}{H^2}\bigg]\Hat{G}_{ab}(uz_1,vz_2) =0\\
     &\bigg[ z^2 \partial^2_{z} - 2z \partial_{z}+\gamma^2 u^4 z_1^4 +\frac{m^2}{H^2}\bigg]\Hat{G}_{ab}(uz_1,vz_2) = -iab H^2 u^2 z_1^2 v^2 z_2^2 \delta (uz_1 - v z_2),
\end{align}
The term \(\gamma^2 u^4 z_1^4\) highlight how the proposal deviates from the standard de Sitter bootstrap.~The above terms must be commuted with the integrals from the seed function~(\ref{Seedgeral}) to derive the differential equation for the seed function. For this work we need the integral formulas up to fourth order, which can be read directly from the general expression:
\begin{align}\label{integralmilagrosa4_2}
&
\int^0_{-\infty} \mathrm{d} z (-z)^l e^{\mathrm{i} \mathrm{a} z} z^n f(u z)= (ia)^n\prod_{j=1}^n (l+1+n-j+u\partial_u)\, \int^0_{-\infty} \mathrm{d} z (-z)^l e^{\mathrm{i} a z} f(u z) 
\end{align}

Then by using Eq.~(\ref{integralmilagrosa4_2}) to commute the time integrals, we have that,
\begin{align}
    \int^0_{-\infty} &dz (-z)^{l} e^{\mathrm{i} \mathrm{a} z}  \big(z^4\big)f(uz)= \big\{ (3+l)(4+l)(1+l)(2+l) + 4(4+l)(2+l)(3+l)u\partial_{u}\nonumber\\
    &+6(3+l)(4+l) u^2\partial_{u}^2+4(l+4)u^3\partial_{u}^3+u^4\partial_{u}^4\big\}\int_{-\infty}^{0}dz(-z)^l e^{iaz} f(uz).
\end{align}
The above equations, when multiplied by $\gamma^2 u^4$ give the desired contribution to the bootstrap equations that come from the ghost particle. Furthermore, the other terms are easily evaluated. Finally, with a little bit of algebra, we can define the following differential operator,
\begin{align}\label{diffdeviation}
    &\mathcal{G}^{l}_{u}\equiv  \mu ^2+\frac{1}{4} (2 l+5)^2-2 \left[(l+2)u-12 \gamma ^2 u^5\right]\partial_{u} + (u^2 + 36u^6\gamma^2)\partial_{u}^2 + u^7 \gamma^2(12 \partial_{u}^3 + u\partial_{u}^4)
\end{align}
Together with the contact term, we have the following bootstrap equations,
\begin{align}
& \mathcal{G}_{u}^{l_1} \mathcal{\hat{I}}_{ \pm \mp}^{l_1 l_2}\left(u, v\right)=0 ,\label{homogeneaghost}\\
& \mathcal{G}_{u}^{l_1} 
\mathcal{\hat{I}}_{ \pm \pm}^{l_1 l_2}\left(u, v\right) =H^2e^{\mp \mathrm{i} (l_1+l_2) \pi / 2} \Gamma\left(5+l_{1}+l_2\right)\left(\frac{u v}{u+v}\right)^{5+l_{1}+l_2} 
.\label{particulaghost}
\end{align}

\bibliographystyle{JHEP}
\bibliography{biblio.bib}

\end{document}